\documentclass[longauth]{aa} % for the long lists of affiliations 
\usepackage{graphicx}
\usepackage{txfonts}
\usepackage{longtable}
\usepackage{float}
\newcommand{\kms}{km\,s$^{-1}$}
\newcommand{\ms}{m\,s$^{-1}$}
\newcommand{\mjup}{M$_{\rm Jup}$}

\newcommand{\lnL}{$\ln \mathcal{L}$}
\newcommand{\aunc}[2]{$^{+#1}_{#2}$}% Uncertainties
\begin{document}

   \title{The CARMENES search for exoplanets around M dwarfs}

   \subtitle{Revisiting the GJ\,317, GJ\,463, and GJ\,3512 systems and two newly discovered planets orbiting GJ\,9773 and GJ\,508.2}

\author{J.\,C.~Morales \inst{1,2}\thanks{Corresponding author: {\tt morales@ice.csic.es}}
        \and
        I.~Ribas \inst{1,2} 
        \and
        S.~Reffert \inst{3}
        \and
        M.~Perger \inst{1,2}
        \and
        S.~Dreizler \inst{4}
        \and
        G.~Anglada-Escud\'e \inst{1,2}
        \and
        V.\,J.\,S.~B\'ejar \inst{5,6}
        \and
        E.~Herrero \inst{2}
        \and
        J.~Kemmer \inst{3}
        \and
        M.~Kuzuhara \inst{7,8}
        \and
        M.~Lafarga \inst{9,10}
        \and
        J.\,H.~Livingston \inst{7,8,11}
        \and
        F.~Murgas \inst{5,6}
        \and
        B.\,B.~Ogunwale \inst{12}
        \and
        L.~Tal-Or \inst{12,4}
        \and
        T.~Trifonov \inst{3,13,14}
        \and
        S.~Vanaverbeke \inst{15,16}
        \and
        P.\,J.~Amado \inst{17}
        \and
        A.~Quirrenbach \inst{3}
        \and
        A.~Reiners \inst{4}
        \and
        J.\,A.~Caballero \inst{18}
        \and
        J.\,F.~Ag\"u\'i Fern\'andez \inst{19}
        \and
        J.~Banegas \inst{20}
        \and
        P.~Chaturvedi \inst{21,22}
        \and
        S.~Dufoer \inst{16}
        \and
        A.\,P.~Hatzes \inst{22}
        \and
        Th.~Henning \inst{13}
        \and
        C.~Rodr\'iguez-L\'opez \inst{17}
        \and
        A.~Schweitzer \inst{23}
        \and
        E.~Solano \inst{18}
        \and
        M.~Zechmeister \inst{4}
        \and
        H.~Harakawa \inst{24}
        \and
        T.~Kotani \inst{7,8,11}
        \and
        M.~Omiya \inst{7,8}
        \and
        B.~Sato \inst{25}
        \and
        M.~Tamura \inst{7,8,26}
        }

 \institute{Institut de Ci\`encies de l'Espai (ICE, CSIC),
        Campus UAB, c/ Can Magrans s/n, 08193 Bellaterra (Barcelona), Spain
        \and
        Institut d'Estudis Espacials de Catalunya (IEEC),
        c/ Esteve Terradas 1, Edifici RDIT, Campus PMT-UPC, 08860 Castelldefels (Barcelona), Spain
        \and
        Landessternwarte, Zentrum f\"ur Astronomie der Universit\"{a}t Heidelberg, K\"{o}nigstuhl 12, 69117 Heidelberg, Germany
        \and
        Institut f\"{u}r Astrophysik und Geophysik, Georg-August-Universit\"{a}t,  Friedrich-Hund-Platz 1, 37077 G\"{o}ttingen, Germany
        \and
        Instituto de Astrof\'isica de Canarias, V\'ia L\'actea s/n, 38205 La Laguna, Tenerife, Spain
        \and
        Departamento de Astrof\'isica, Universidad de La Laguna, 38206 La Laguna, Tenerife, Spain
        \and
        Astrobiology Center, 2-21-1 Osawa, Mitaka, Tokyo 181-8588, Japan
        \and
        National Astronomical Observatory of Japan, 2-21-1 Osawa, Mitaka, Tokyo 181-8588, Japan
        \and
        Department of Physics, University of Warwick, Gibbet Hill Road, Coventry CV4 7AL, UK
        \and
        Centre for Exoplanets and Habitability, University of Warwick, Coventry CV4 7AL, UK
        \and
        Astronomical Science Program, Graduate University for Advanced Studies (SOKENDAI), 2-21-1, Osawa, Mitaka, Tokyo, 181-8588, Japan
        \and
        Department of Physics, Ariel University, Ariel 40700, Israel
        \and
        Max-Planck-Institut für Astronomie, Königstuhl 17, 69117 Heidelberg, Germany
        \and
        Department of Astronomy, Faculty of Physics, Sofia University “St. Kliment Ohridski”, 5 James Bourchier Blvd., 1164 Sofia, Bulgaria
        \and
        Centre for mathematical Plasma-Astrophysics, Department of Mathematics, KU Leuven, Celestijnenlaan 200B, 3001 Heverlee, Belgium
        \and
        Vereniging Voor Sterrenkunde, Oude Bleken 12, 2400 Mol, Belgium
        \and
        Instituto de Astrof\'isica de Andaluc\'ia (IAA-CSIC), Glorieta de la Astronom\'ia s/n, 18008 Granada, Spain
        \and
        Centro de Astrobiolog\'ia (CSIC-INTA), ESAC Campus, Camino Bajo del Castillo s/n, 28692 Villanueva de la Ca\~nada, Madrid, Spain
        \and
        Centro Astron\'omico Hispano en Andaluc\'ia (CAHA), Observatorio de Calar Alto, Sierra de los Filabres, 04550 G\'ergal, Almer\'ia, Spain
        \and
        Departamento de F\'{i}sica de la Tierra y Astrof\'{i}sica and IPARCOS-UCM (Instituto de F\'{i}sica de Part\'{i}culas y del Cosmos de la UCM), Facultad de Ciencias F\'{i}sicas, Universidad Complutense de Madrid, 28040, Madrid, Spain
        \and
        Department of Astronomy and Astrophysics, Tata Institute of Fundamental Research, 400005 Mumbai, India
        \and
        Th\"uringer Landessternwarte Tautenburg, Sternwarte 5, 07778 Tautenburg, Germany
        \and
        Hamburger Sternwarte, Universit\"at Hamburg, Gojenbergsweg 112, 21029 Hamburg, Germany
        \and
        Subaru Telescope, National Astronomical Observatory of Japan, 650 North A`oh$\bar{o}$k$\bar{u}$ Place, Hilo, HI  96720, USA
        \and
        Department of Earth and Planetary Sciences, School of Science,  Institute of Science Tokyo, 2-12-1 Ookayama, Meguro-ku, Tokyo 152-8551, Japan
        \and
        Department of Astronomy, Graduate School of Science, The University of Tokyo, 7-3-1, Hongo, Bunkyo-ku, Tokyo, 113-0033, Japan
        }

   \date{Received 1 April 2025 / Accepted 30 June 2025}

% \abstract{}{}{}{}{} 
% 5 {} token are mandatory
 
  \abstract
   {Surveys for exoplanets indicate that the occurrence rate of gas giant planets orbiting late-type stars in orbits with periods shorter than 1000 days is lower than in the case of Sun-like stars. This is in agreement with planet formation models based on the core or pebble accretion paradigm.}
   {The CARMENES exoplanet survey has been conducting radial-velocity observations of several targets that show long-period trends or modulations that are consistent with the presence of giant planets at large orbital separations.}
   {We present an analysis of five such systems that were monitored with the CARMENES spectrograph, as well as with the IRD spectrograph. In addition, we used archival data to improve the orbital parameters of the planetary systems.}
   {We improve the parameters of three previously known planets orbiting the M dwarfs GJ\,317, GJ\,463, and GJ\,3512. We also determine the orbital parameters and minimum mass of the planet GJ\,3512\,c, for which only lower limits had been given previously. Furthermore, we present the discovery of two new giant planets orbiting the stars GJ\,9733 and GJ\,508.2, although for the second one only lower limits to the orbital properties can be determined.}
   {The new planet discoveries add to the short list of known giant planets orbiting M-dwarf stars with subsolar metallicity at long orbital periods above 2000 days. These results reveal that giant planets appear to form more frequently in wide orbits than in close-in orbits around low-mass and lower metallicity stars.}
   
   \keywords{planetary systems -- techniques: radial velocities -- stars: late-type -- stars: individual (GJ\,3512, GJ\,317, GJ\,463, GJ\,9773, GJ\,508.2) -- planets and satellites: detection -- planets and satellites: formation}

   \maketitle
%
%-------------------------------------------------------------------

\section{Introduction}

After three decades of discoveries, the present sample of about 6000 exoplanet candidates has greatly contributed to the improvement of planet formation models by providing a rich variety of observational constraints. Current models yield a description of the statistical occurrence of exoplanets as a function of their masses and orbital properties that results in  a good match to the observed distributions \citep{Emsenhuber2021,Schlecker2022}. However, there are still areas of the parameter space where model predictions and observations are at odds. One such regime is giant planets orbiting M dwarfs, that is, stars with masses between approximately 0.08\,M$_\odot$ and 0.6\,M$_\odot$ \citep{Baraffe1996,Mann2015,Benedict2016,Schweitzer2019}. 

Because of the correlation between stellar mass, protoplanetary disc mass, and planet mass \citep{Pascucci2016,Ansdell2017,Manara2018,Tychoniec2020,AlmendrosAbad2024}, giant planets orbiting M dwarfs are expected to be much scarcer  than those orbiting solar-type stars. In spite of the favourable detection bias, current databases, such as the NASA Exoplanet Archive\footnote{ \url{https://exoplanetarchive.ipac.caltech.edu/}, accessed on 20 May 2025.} , show that only 168 giant planet candidates with $m_p$ above 0.2 Jupiter masses (\mjup) have been reported around M dwarfs, while over 1000 such planets are known to orbit FGK stars ($0.6\,{\rm M}_\odot < M_\star < 1.6\,{\rm M}_\odot$). About two thirds of the M-dwarf giant planet discoveries have been performed with the microlensing technique, but the radial velocity (hereafter, RV) technique and, more recently, transit searches have also contributed a score of planets each. The microlensing technique yields valuable statistical data \citep{Gould2010,Cassan2012,Suzuki2016}, but the resulting planet parameters typically have large error bars, because this technique is primarily sensitive to the mass ratio between the planet and the host star, with the mass of the star being usually inferred from models with considerable uncertainty, and only limited information is available on the orbital properties.

Of particular interest to constrain models is the region beyond the so-called `snow line' (the distance from the star beyond which water condenses into the solid phase), which is the location beyond which giant planets are expected to form \citep[e.g.][]{Ida2005}. When we restrict the parameters to orbital semi-major axis values greater than the present-day ice line \citep[following the scaling relation in][$a_{\rm ice} \approx 2.7 (L/L_{\odot})^{1/2}$, but ignoring the luminosity evolution in the pre-main-sequence phase]{Ida2005}, the number of known M-dwarf giant planets is approximately 100; of them, three quarters were discovered via microlensing and only about 15 from RV surveys. Therefore, the statistical knowledge of the occurrence and properties of giant planets around low-mass stars in cool orbits is still very limited. For comparison, about 400 giant planets orbiting beyond the present-day ice line of FGK stars have been found, the majority of them using the RV technique because of the low transit probability at large separations.

Planets with orbits beyond the snow line have typical periods of hundreds to thousands of days, which implies that RV surveys need to accumulate data over several years to reveal their presence in an unambiguous way. While systematic RV searches for solar-type stars started over three decades ago, this is not the case for M dwarfs. The CARMENES\footnote{Calar Alto high-Resolution search for M dwarfs with Exoearths with
Near-infrared and optical \'Echelle Spectrographs; \url{https://carmenes.caha.es/}} survey is one such survey that is focused on the search for exoplanets around late-type stars \citep{Quirrenbach2018}. It started in 2016 and has now been operating for over nine years. This has made data available with a time baseline that is sufficiently long to sample the orbital regions beyond the snow line of the over 350 M dwarfs being monitored \citep{Ribas2023}. The surveys performed with the High Accuracy Radial velocity Planet Searcher spectrograph \citep[hereafter HARPS,][]{Mayor2003} and the High Resolution Echelle Spectrometer \citep[hereafter HIRES,][]{Vogt1994}, which started more than 20 years ago, have also reported several long-period candidate systems around early and mid M-dwarf stars \citep[e.g][]{Butler2006,Johnson2007,Moutou2015}.

In this paper, we reanalyse data on three M-dwarf stars that host giant planets that are being observed within the context of the CARMENES survey. These are GJ\,317 \citep{Johnson2007}, GJ\,463 \citep{Endl2022}, and GJ\,3512 \citep{Morales2019}. Furthermore, we announce two new long-period giant planet candidates orbiting the early- and mid-M-type stars GJ\,508.2 and GJ\,9773\footnote{The GJ numbers higher than 9000 were not assigned by \cite{Gliese1957} or \cite{GlieseJahreiss1979,GlieseJahreiss1991}, but for homogeneity we use the `GJ' nomenclature of \cite{Woolley1970} for Wolf~1014.}, respectively. In Sect.\,\ref{sec:host_stars} we summarise the properties of these stars. Section\,\ref{sec:observations} describes the observations used to characterise the properties of the systems. We study the spectroscopic indices and the photometry of the host stars in Sect.\,\ref{sec:activityAnalysis} to infer their rotation periods. The analysis of RV data is described in Sect.\,\ref{sec:RVanalysis}, and in Sect.\,\ref{sec:astrometry} we combine the RV orbits with constraints from astrometry to obtain absolute masses of the planets. The results are discussed in Sect.\,\ref{sec:discussion} and conclusions are summarised in Sect.\,\ref{sec:conclusions}.

%--------------------------------------------------------------------
\section{Host star properties}
\label{sec:host_stars}

\begin{table*}[t]
\centering
\caption[]{Properties of the stars GJ\,3512, GJ\,317, GJ\,463, GJ\,508.2, and GJ\,9773.}
\label{tab:stars}
\begin{tabular}{lcccccr}
\hline
\hline
\noalign{\smallskip}
             & GJ\,3512             & GJ\,317             & GJ\,463             & GJ\,9773            & GJ\,508.2            & Ref. \\
\noalign{\smallskip}
\hline                                       \noalign{\smallskip}

Karmn        & J08413+594           & J08409-234          & J12230+640          & J22125+085          & J13209+342          & \\
Spectral type& M5.5\,V              & M3.5\,V                & M3.0\,V                & M3.0\,V                & M1.0\,V                & 1 \\
$\alpha$ (J2000)& 08:41:20.13       & 08:40:59.21         & 12:23:00.16         & 22:12:35.94         & 13:20:58.05         & 2 \\
$\delta$ (J2000)& +59:29:50.44      & --23:27:22.59       & +64:01:50.96        & +08:32:55.50        & +34:16:44.20        & 2 \\
pm$\alpha$ (mas\,a$^{-1}$) & --260.276$\pm$0.023  & --461.344$\pm$0.041 & --673.116$\pm$0.018 & 106.380$\pm$0.026   & 483.884$\pm$0.014   & 2 \\
pm$\delta$ (mas\,a$^{-1}$) & --1279.562$\pm$0.026 & 805.221$\pm$0.038   & 373.311$\pm$0.021   & --671.154$\pm$0.041 & --297.069$\pm$0.016 & 2 \\
$\pi$ (mas)  & 105.294$\pm$0.031    & 65.877$\pm$0.041    & 54.447$\pm$0.019    & 62.878$\pm$0.022    & 60.029$\pm$0.019    & 2  \\
$d$ (pc)     & 9.4973$\pm$0.0028    & 15.1798$\pm$0.0094  & 18.3664$\pm$0.0066  & 15.9037$\pm$0.0056  & 16.6585$\pm$0.0054  & \\

\hline                                       
\noalign{\smallskip}
\multicolumn{7}{c}{Photometry}\\
\noalign{\smallskip}                         
\hline
\noalign{\smallskip}
$BP$ (mag)         &15.3833$\pm$0.0046    & 12.2327$\pm$0.0033  & 11.8252$\pm$0.0030  & 12.2297$\pm$0.0033  & 10.8363$\pm$0.0029  & 2 \\
$V$ (mag)          &15.166$\pm$0.072$^{(a)}$& 11.98$\pm$0.04    & 11.59$\pm$0.06      & 11.96$\pm$0.01      & 10.61$\pm$0.01      & 3 \\
$G$ (mag)          &13.0823$\pm$0.0029    & 10.7651$\pm$0.0028  & 10.5524$\pm$0.0028  & 10.9082$\pm$0.0028  & 9.7727$\pm$0.0028   & 2 \\
TESS (mag)        &11.576$\pm$0.008      & 9.511$\pm$0.007     & 9.394$\pm$0.007     & 9.734$\pm$0.007     & 8.97$\pm$0.14       & 4 \\  
$RP$ (mag)               &11.7121$\pm$0.0041    & 9.5689$\pm$0.0039   & 9.4291$\pm$0.0038   & 9.7698$\pm$0.0038   & 8.7456$\pm$0.0038   & 2 \\
$J$ (mag)          &9.615$\pm$0.023       & 7.934$\pm$0.027     & 7.937$\pm$0.029     & 8.277$\pm$0.021     & 7.398$\pm$0.022     & 5 \\
$H$ (mag)          &8.996$\pm$0.028       & 7.321$\pm$0.071     & 7.343$\pm$0.020     & 7.681$\pm$0.024     & 6.793$\pm$0.018     & 5 \\
$K$ (mag)          &8.668$\pm$0.021       & 7.028$\pm$0.020     & 7.122$\pm$0.018     & 7.472$\pm$0.027     & --                  & 5 \\
$W1$ (mag)               &8.430$\pm$0.023       & 6.867$\pm$0.056     & 6.962$\pm$0.051     & 7.239$\pm$0.035     & 6.516$\pm$0.069     & 6 \\
$W2$ (mag)         &8.215$\pm$0.019       & 6.766$\pm$0.021     & 6.883$\pm$0.020     & 7.147$\pm$0.019     & 6.371$\pm$0.023     & 6 \\
$W3$ (mag)         &8.054$\pm$0.020       & 6.704$\pm$0.016     & 6.815$\pm$0.017     & 7.064$\pm$0.017     & 6.363$\pm$0.016     & 6 \\
$W4$ (mag)         &8.051$\pm$0.199       & 6.502$\pm$0.059     & 6.634$\pm$0.057     & 6.918$\pm$0.103     & 6.237$\pm$0.045     & 6 \\

\noalign{\smallskip}                         
\hline                                       
\noalign{\smallskip}                         
\multicolumn{7}{c}{\textit{Gaia} DR3 parameters}\\    \noalign{\smallskip}                         
\hline
\noalign{\smallskip}
 Excess noise (mas) & 0.264          & 0.358               & 0.183               & 0.133               & 0.114               & 2 \\
Significance       & 75.4           & 174.7               & 46.0                & 23.4                & 16.8                & 2 \\
\hline                                                                                                                    
\noalign{\smallskip} 
\multicolumn{7}{c}{Stellar properties}\\                     \noalign{\smallskip}                                         
\hline                                                       
\noalign{\smallskip}
$M$ (M$_{\odot}$) & 0.123$\pm$0.009 & 0.402$\pm$0.014     & 0.486$\pm$0.014     & 0.358$\pm$0.011     & 0.54$\pm$0.12       & 7 \\
$R$ (R$_{\odot}$) & 0.139$\pm$0.005 & 0.403$\pm$0.009     & 0.483$\pm$0.008     & 0.362$\pm$0.005     & 0.53$\pm$0.12       & 7 \\
$T_{\rm eff}$ (K) & 3141$\pm$38     & 3496$\pm$37         & 3563$\pm$28         & 3528$\pm$25         & 3720$\pm$12         & 8 \\

[Fe/H] (dex)      & --0.07$\pm$0.16 & --0.02$\pm$0.08     & --0.16$\pm$0.10     & --0.20$\pm$0.09     & --0.15$\pm$0.04     & 8 \\
$v\sin i$ (\kms)  & $<$2            & $<$2.5$^{(b)}$      & $<$2                & $<$2.0              & $<$2.0              & 9 \\
pEWH$\alpha$ (\AA)&--1.339$\pm$0.013$^{(c)}$& 0.202$\pm$0.024& 0.322$\pm$0.013  & 0.225$\pm$0.012     & 0.379$\pm$0.009     & 10 \\
$P_{\rm rot}$ (d) & 83              & 57.5$\pm$1.7        & 32.9$\pm$1.1        & 56$\pm$8$^{(d)}$                  & 36$\pm$6$^{(d)}$    & 11 \\
\hline
\end{tabular}
\tablebib{
(1)~\cite{Reid1995}; (2)~\cite{Gaia2023}; (3)~UCAC4; (4)~TESS; (5)~2MASS; (6)~WISE; (7)~\cite{Schweitzer2019}; (8)~\cite{Marfil2021}; (9)~\cite{Reiners2018}; (10)~\cite{Fuhrmeister2020}; (11)~\cite{Shan2024}.\\
\textbf{Notes:} $^{(a)}$ value from \cite{APASS9}, $^{(b)}$ value from \cite{Browning2010}, $^{(c)}$ value from \cite{Schofer2019}, $^{(d)}$ values from this work.}
\end{table*}

Table\,\ref{tab:stars} summarises the properties of the host stars analysed here, which are compiled in the Carmencita catalogue \citep{Caballero2016}. We compute the distance to the host star from \textit{Gaia} parallax measurements \citep{Gaia2016, Gaia2020}, and we also retrieve from \textit{Gaia} the astrometric excess noise and its significance. The stellar atmospheric parameters ($T_{\rm eff}$, $\log{g}$, and [Fe/H]) were derived with the {\tt SteParSyn}\footnote{ \url{https://github.com/hmtabernero/SteParSyn/}} code \citep{Tabernero2022} using the line list and model grid described by \cite{Marfil2021} and comparing \ion{Fe}{i} and \ion{Ti}{i} and \ion{TiO}{} spectral features with synthetic models. Their radii and masses are determined as described in \cite{Schweitzer2019}, making use of the bolometric luminosity measured from several broad band apparent magnitudes \citep{Cifuentes2020} and empirical mass-radius relationships \citep{Schweitzer2019}.

GJ\,3512 is an M5.5 spectral type star with a mass of 0.123\,M$_{\odot}$. \cite{Morales2019} reported the detection of a giant planet with a mass of $\sim$0.46\,\mjup\, orbiting the star in a highly eccentric orbit with a period of $\sim$204 days by means of the RV method using the CARMENES high-resolution spectrograph. In that work, a rotation period of $\sim$87\,days was identified from the differential line width index \citep[dLW,][]{Zechmeister2018} and photometric data, which is consistent with the period of $\sim$83\,days based on MEarth data \citep{Pass2023}. The residuals of the fit to the RV time series indicated also a long-term trend. For this reason, spectroscopic observations of the system were continued within the CARMENES exoplanet survey. We present here the properties of a second giant planet in the system with a longer orbital period of about 9.3 years.

GJ\,317 is an M3.5 spectral type star with a mass of 0.402\,M$_{\odot}$ as determined from the CARMENES spectra analysed in this work. A $\sim$1.2\,\mjup\, minimum mass planet orbiting this system every $\sim693$\,days was reported by \cite{Johnson2007} from observations with the high-resolution echelle spectrograph HIRES \citep{Vogt1994} mounted on the Keck I 10\,m telescope. The residuals of the one-planet model were found to be consistent with the presence of an additional Keplerian orbital signal in the system at longer orbital period. Therefore, the target was further observed with the HARPS spectrograph \citep{Mayor2003} at the 3.6\,m telescope at La Silla Observatory, and the Planet Finder Spectrograph \citep[hereafter PFS,][]{Crane2006} on the Magellan telescope at Las Campanas Observatory. A second outer planet with a minimum mass of $\sim$1.6\,\mjup\, and an orbital period of $\sim$18.5\, years was confirmed \citep{AngladaEscude2012,Feng2020,Rosenthal2021}. The analysis of photometric time series yielded a rotation period of the star of 57.50\,days \citep{Irving2023}. We have gathered 41 more observations for this target with CARMENES, increasing the time baseline of the available observations by 6 years to better determine the properties of the outer orbit. 

GJ\,463 is an M3 spectral type star with a mass of 0.486\,M$_{\odot}$. A linear trend was detected in the data early in the CARMENES survey \citep{Ribas2023}, and radial velocity monitoring was performed to investigate the nature of the signal. Before we could clearly identify the orbital period with our observations, a giant planet with an orbital period $\sim9.4$\,years was announced by \cite{Endl2022} using the High Resolution Spectrograph \citep[hereafter HRS,][]{Tull1998} at the Hobby-Eberly Telescope \citep[HET,][]{Endl2003} and HIRES spectrograph at the Keck telescope \citep{Vogt1994}. Subsequently, an absolute mass of the planet of 3.6$\pm$0.4\,\mjup\, was inferred by \cite{Sozzetti2023} from the analysis of \textit{Hipparcos} and \textit{Gaia} proper motion anomalies \citep{Kervella2022}. A rotation period of 32.9\,days has been proposed from photometric variability for this system \citep[FAP$<$0.1\%,][]{DiezAlonso2019}, although a longer value of 51.5\,days is also suggested by the analysis of the Ca IRT lines \citep{Shan2024} in the CARMENES observations. We add a total of 160 RVs determined using the CARMENES visual channel spectrograph to the analysis of this system.

Finally, GJ\,9773 and GJ\,508.2 are two M dwarfs of spectral type M3.0 and M1.0 and masses of 0.358\,M$_{\odot}$ and 0.54\,M$_{\odot}$, respectively \citep{Schweitzer2019}.
Both stars were observed with the high resolution CAFE and FEROS spectrographs \citep{Jeffers2018} to ascertain that they are not them being spectroscopic binaries. Besides, close companions were not found using the FastCam lucky imager and adaptive optics instrumentation \citep{Jodar2013,WardDuong2015}. A rotation period of 43.53\,days is suggested from photometric monitoring for GJ\,508.2 \citep{Oelkers2018}, while there is no reported value in the literature for GJ\,9773.

\section{Observations}
\label{sec:observations}
\subsection{CARMENES spectroscopic data}
\label{subsec:CARMENESdata}

GJ\,3512, GJ\,317, GJ\,463, GJ\,9773, and GJ\,508.2 were spectroscopically followed-up as part of the CARMENES exoplanets survey \citep{Ribas2023}. The CARMENES instrument \citep{Quirrenbach2018} consists of two high-resolution spectrographs, one covering the visible wavelengths, from 520 to 960 nm (VIS), and one in the near-infrared, covering from 960 to 1710 nm (NIR) with resolving powers of 94600 and 80400, respectively. Observations were taken between December 2016 and February 2024. The median signal-to-noise ratio (S/N) of the observations of GJ\,3512 at the central orders of the VIS and NIR channels are $\sim$35 and $\sim$70, respectively. For other, brighter targets, the S/N ranges between 80 and 130, and 90 and 150 in the VIS and NIR channels, respectively. Dense sampling was secured at the start of the survey for GJ\,3512, GJ\,463 and GJ\,9773, but the spectroscopic sampling rate was reduced after the observation of long period signals. In the case of GJ\,317, monthly observations were scheduled from December 2017 to characterise the long period candidate GJ\,317\,c announced in \cite{AngladaEscude2012}.

The CARMENES spectra were processed with the CARMENES pipeline {\tt caracal}\footnote{CARMENES Reduction And CALibration.} \citep{Caballero2016} and the radial velocities were computed using {\tt serval}\footnote{SpEctrum Radial Velocity AnaLyser. \url{https://github.com/mzechmeister/serval}} \citep{Zechmeister2018}, which employs the template matching technique. The computed RVs are corrected for nightly zero-points, calculated using a statistical sample of stable stars in the survey, and by instrumental drift, and secular acceleration \citep[see][for further details]{Ribas2023}. Clear RV outliers were removed using a 3$\sigma$ filter from the median value. Measurements with uncertainties above 3$\sigma$ the median value were also discarded. Prior to November 2016, the original design of the thermal preparation unit did not provide the stability needed to achieve the required radial velocity precision. An intervention in the instrument at that time (November 2016; BJD~$\approx$~2457700) enabled scientific use of the NIR RVs, although without fully meeting those requirements \citep{Bauer2020,Varas2025}. A major upgrade was later carried out between February 2021 (BJD~$\approx$~2459250) and 1 June 2022 (BJD~$=$~2459731) under the CARMENES-PLUS project, which introduced stricter radial velocity precision requirements. We inspected the NIR RV values taken between these dates for the systems analysed here (a total of 32 measurements), which showed a similar dispersion to previous values. For these reason, we decided not to remove them from the analysis. Table\,\ref{tab:obslog} provides the observation log and number of RVs used in this work. The individual RV measurements are listed in Tables \ref{tab:RVs_GJ3512} to \ref{tab:RVs_GJ508}.

\subsection{IRD spectroscopic data}
\label{subsec:IRDdata}
GJ\,3512 was also monitored to measure its RV using the InfraRed Doppler (IRD; Tamura et al. 2012; Kotani et al. 2018) instrument, a high-resolution (R~$\sim$ 70,000) near-infrared (950 – 1730 nm) spectrograph on the Subaru 8.2\,m telescope. The IRD observations were conducted as part of the Subaru Strategic Program (SSP), which aims at searching for exoplanets around mid-to-late M dwarfs. A total of 39 spectra of GJ\,3512 were taken between October 2019 and November 2022 on 24 different nights.
 
We used a laser-frequency comb (LFC) to calibrate instrumental wavelength drifts. The LFC light was injected into the IRD spectrograph through a multi-mode fibre  simultaneously with the light from GJ\,3512. We reduced the raw IRD data following the standard procedures \citep{Kuzuhara2018,Hirano2020}, producing one-dimensional stellar spectra with a typical S/N of 50–90 pixel$^{-1}$ at 1000 nm. To estimate GJ\,3512 RVs, we applied the forward modelling technique developed by \citep{Hirano2020} to the one-dimensional spectra.

As reported in \cite{Gorrini2023}, zero-point offsets are present in IRD RV measurements obtained at each observing sequence that consists of bright nights. We corrected for those offsets using the post-processing technique adopted in \cite{Gorrini2023}, which removed only instrumental RV signals that are common among multiple IRD targets observed in each sequence. Because the $YJ$ and $H$ bands are covered by two different detectors \citep{Kotani2018}, we corrected the zero-point offsets separately for the $YJ$ and $H$ bands. In addition to these corrections, we implemented an additional step to mitigate the effects of inaccurate RV measurements, as performed in \cite{Kuzuhara2024} for RV measurements of the star Gliese\,12. The IRD RV pipeline divides a spectrum into many segments, individually calculates RVs for each segment, and then integrates them into a single RV value for the spectrum. At the final integration step, we statistically removed outlier segments. As a result, we measured the stellar RVs, with typical internal errors of 3–5 m\,s$^{-1}$.

\begin{table}
\caption[]{Summary of spectroscopic data.}
\label{tab:obslog}
\begin{tabular}{@{}rc@{}c@{}c@{}r@{}}
\hline
\hline
           &      & \#          &  Time span & \\
Instrument & Date & Spectra &  [d]           & Source\\
\hline
\noalign{\smallskip}
\multicolumn{4}{c}{GJ\,3512}\\
\noalign{\smallskip}
\hline
\noalign{\smallskip}
CARM-VIS   & Dec. 2016 -- Nov. 2024  & 221 & 2869 & TW \\
CARM-NIR   & Dec. 2016 -- Nov. 2024  & 225 & 2869 & TW \\
IRD        & Oct. 2019 -- Nov. 2022  &  39 & 999 & TW \\
\hline
\noalign{\smallskip}
\multicolumn{4}{c}{GJ\,317}\\
\noalign{\smallskip}
\hline
\noalign{\smallskip}
HIRES      & Jan. 2000 -- Dec. 2013 &  63 & 5086 & TO19 \\
HARPS      & Jan. 2010 -- Apr. 2018 &  132 & 2991 & T20 \\
PFS        & Feb. 2012 -- Dec. 2018 &  32 & 2520 & F20 \\
CARM-VIS   & Dec. 2017 -- Jan. 2025 &  41 & 2595 & TW  \\
CARM-NIR   & Dec. 2017 -- Jan. 2025 &  38 & 2595 & TW  \\
\hline
\noalign{\smallskip}
\multicolumn{4}{c}{GJ\,463}\\
\noalign{\smallskip}
\hline
\noalign{\smallskip}
HRS        & Apr. 2008 -- Jun. 2013 &  53 & 1875 & E22 \\
HIRES      & Jan. 2010 -- Jan. 2022 &  17 & 4377 & E22 \\
CARM-VIS   & Mar. 2016 -- Feb. 2025 & 160 & 3282 & TW \\
CARM-NIR   & Mar. 2017 -- Feb. 2025 & 146 & 2909 & TW \\
\hline
\noalign{\smallskip}
\multicolumn{4}{c}{GJ\,9773}\\
\noalign{\smallskip}
\hline
\noalign{\smallskip}
CARM-VIS   & Jul. 2016 -- Nov. 2024 & 138 & 3050 & TW \\
CARM-NIR   & Jul. 2017 -- Nov. 2024 & 131 & 2677 & TW \\
\hline
\noalign{\smallskip}
\multicolumn{4}{c}{GJ\,508.2}\\
\noalign{\smallskip}
\hline
\noalign{\smallskip}
CARM-VIS   & Feb. 2016 -- Feb. 2025 &  26 & 3287 & TW \\
CARM-NIR   & Mar. 2017 -- Feb. 2025 &  20 & 2908 & TW \\
\hline
\end{tabular}
\tablebib{
(E22)~\cite{Endl2022}; (F20)~\cite{Feng2020}; (T20)~\cite{Trifonov2020}; (TO19)~\cite{TalOr2019}; (TW)~this work.}
\end{table}

\subsection{Bibliographic spectroscopic data}
\label{subsec:LiteratureData}
We made use of the RVs measured for GJ\,463 and GJ\,317 using the HARPS \citep{Mayor2003}, HIRES \citep{Vogt1994}, HRS \citep{Tull1998}, and PFS \citep{Crane2006}. We employed the nightly zero-point corrected HARPS and HIRES RV data from \cite{Trifonov2020} and \cite{TalOr2019}, respectively. The RVs determined from PFS spectra were taken from \cite{Feng2020}. Table\,\ref{tab:obslog} summarises the properties of the different RV sets utilised in this work.

\subsection{Photometric data}
\label{subsec:photometricData}

In order to evaluate the variability of the new exoplanet candidate hosting stars GJ\,9773 and GJ\,508.2, we photometrically monitored them with different facilities. GJ\,9773 was observed with the 4k$\times$4k back-illuminated CCD imaging camera LAIA mounted on the 0.8\,m Joan Or\'o Telescope (TJO) at the Montsec Observatory (Sant Esteve de la Sarga, Spain) between April and July 2024. $R$-band images were calibrated using the usual dark, bias and flat field frames by means of the ICAT pipeline \citep{Colome2006}. Differential photometry was obtained making use of {\tt AstroImageJ} \citep{Collins2017}. Appropriate apertures and comparison stars were defined to minimise the dispersion of the photometry. A total of 75 epochs on 21 nights were obtained. GJ\,508.2 was also observed from the TJO at Montsec Observatory with the same instrumental setup between July 2024 and July 2024. A total of 539 epochs were obtained over 87 nights.

We also obtained photometry of GJ\,9773 with the 0.9\,m Ritchey-Chr\'etien telescope at the Observatorio de Sierra Nevada (OSN, Granada, Spain). The telescope is equipped with a CCD camera Andor Ikon-L DZ936N-BEX2-DD 2k$\times$2k having a field of view of 13.2$\times$13.2 arcmin. The camera is based on a back-illuminated CCD chip, with high quantum efficiency from the ultraviolet to the near-IR, and has a thermo-electrical cooling system down to $-$100\,$^{\circ}$C for negligible dark current. Our set of observations, collected in the Johnson $V$ and $R$ bands, consists of $\sim$1500 epochs over 78 nights between June and December 2024. Typically, 20 exposures where obtained in the $V$ and $R$ filters per night, with and exposure time of 50\,s and 30\,s, respectively. All CCD measurements were obtained by the method of aperture photometry. Each CCD frame was corrected in a standard way for bias and flat field. Different aperture sizes were tested to select the best ones for our observations. A number of nearby and relatively bright stars within the frames were used as reference stars.

We observed GJ\,508.2 in the $V$ band from February to December 2024 using the 0.4\,m telescopes of Las Cumbres Observatory Global Telescope \citep[LCOGT;][]{Brown2013} at the McDonald (Texas, USA) and the Teide (Spain) observatories. GJ\,9773 was also observed in $V$ band using LCOGT facilities at the same sites and also the Cerro Tololo Inter-American Observatory (CTIO) in Chile and the South African Astronomical Observatory (SAAO) from September to December 2024. The 0.4\,m telescopes have a 9.5k$\times$6.4k QHY600 CMOS camera with a pixel scale of 0.74\,arcsec providing a field of view of 30$\times$30\,arcmin in readout mode "central". On each night, typically 10 individual exposures with an exposure time of 40\,s were taken. Furthermore, GJ\,508.2 and GJ\,9773 photometry was also gathered using the 1\,m telescopes at the same LCOGT sites as well as at the Siding Spring Observatory in Australia between May and December 2024. The 1\,m telescopes are equiped with 4k$\times$4k SINISTRO cameras with a pixel scale of 0.389\,arcsec providing a field of view of 26.5$\times$26.5\,arcmin. In this case observations were carried out with the Johnson-Cousins/Bessell $B$ filter, using an exposure time of 10 and 25 seconds, for GJ\,508.2 and GJ\,9773 respectively, and a total of 3 frames every observing night. LCOGT raw data were processed using the {\tt BANZAI} pipeline \citep{McCully2018}, which includes bad pixel, bias, dark and flat field corrections for each individual night. The images were aligned, and we performed aperture photometry of the star and several reference stars, and obtained the relative differential photometry using {\tt AstroImageJ} \citep{Collins2017}, adopting an optimised aperture. The flux values obtained for all stars where median-normalised. 

We also gathered photometric data for both stars from the remote telescope hosting facility e-EYE (Entre Encinas y Estrellas) located at Fregenal de la Sierra in Badajoz, Spain\footnote{\url{www.e-eye.es/}}. We used a 16-inch ODK corrected-Dall-Kirkham reflector and collected observations with a Kodak KAF-16803 CCD chip mounted on ASA DDM85. The images were reduced and differential photometry was performed using the {\tt LesvePhotometry} package\footnote{\url{www.dppobservatory.net}}. We obtained  a total of 179 epochs in the $V$ and $R$ bands for GJ\,9773 from January 2024 to February 2025 and 265 $R$ band observations for GJ\,508.2 from June 2023 to February 2025, respectively.

\section{Stellar activity analysis}
\label{sec:activityAnalysis}
\subsection{Analysis of spectroscopic indices}
\label{sec:activity_spectra}
\begin{figure}[t]
\centering
\includegraphics[width=\columnwidth]{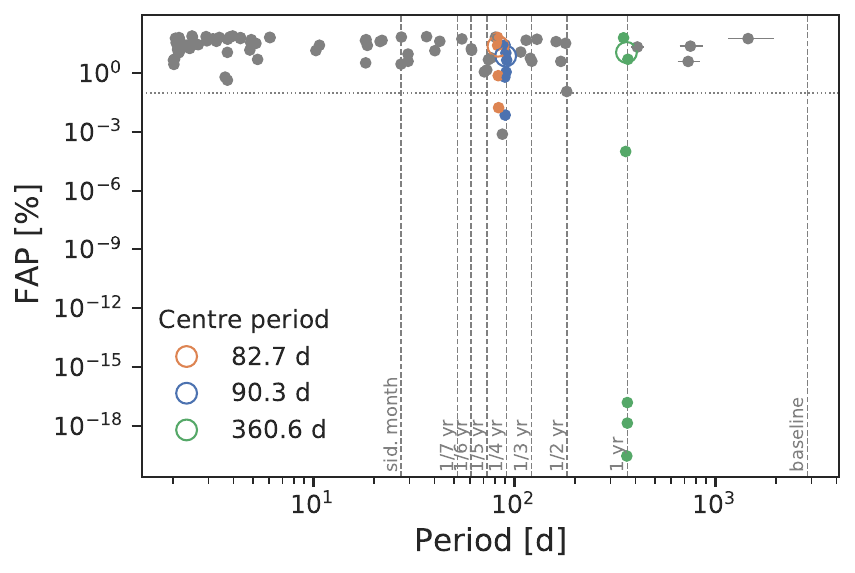}
\includegraphics[width=\columnwidth]{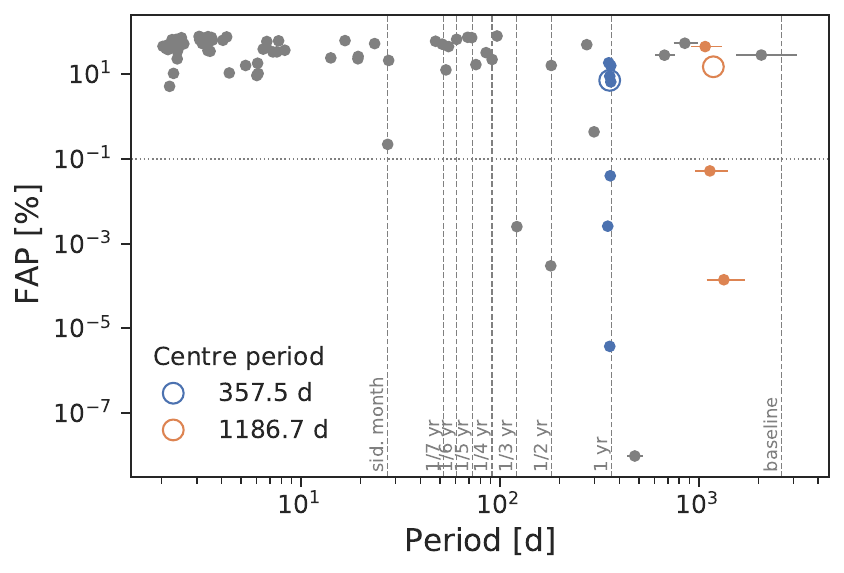}
\caption{Clustering periodogram analysis \citep{Kemmer2025} for the activity indices of GJ\,3512 (top) and GJ\,9773 (bottom). The 0.1\% FAP threshold is shown as a horizontal dotted grey line, while vertical dashed lines mark the position of common periods caused by the sampling of data and their harmonics as labelled. Symbols with different colours correspond to clusters of periods that appear on different activity indicators. The mean values are reported in the legend.}
\label{fig:clustering}
\end{figure}

As stated in Sect. \ref{sec:host_stars} and listed in Table\,\ref{tab:stars}, the targets studied in this work have reported rotation periods well above 10 days and, therefore, they are expected to display mild or low levels of activity. Only the latest type star in the sample, GJ\,3512, shows the H$\alpha$ line in emission with a pseudo-equivalent width of $-1.339 \pm 0.013$~\AA~\citep{Schofer2019}, which indicates the presence of chromospheric activity. \cite{Morales2019} reported GJ\,3512 to be a moderately active star with a rotation period of 87$\pm$5\,days derived from photometric data and spectroscopic indices. This is consistent with the more recent determination by \cite{Pass2023}. As a further check, we inspected the spectroscopic activity indicators in search for common periodicities by taking advantage of the longer time baseline now available from CARMENES observations. We followed the methodology described by \cite{Kemmer2025}. This consists in the computation of the Generalised Lomb-Scargle periodogram \citep[heareafter GLS,][]{Zechmeister2009} for all activity indicators derived from CARMENES spectra, pre-whitening the identified signals until reaching a conservative false-alarm-probability (FAP) of 80\%. Then, the periodograms are scanned for common periods among the different indicators using the {\tt DBSCAN} clustering algorithm, which counts the number of periods overlapping within a distance of half the peak width in the frequency space, and identifies clusters as those with at least the 3 peaks from different activity indicators. The top panel in Figure\,\ref{fig:clustering} shows that, in addition to the cluster of periods around 1 year most probably caused by observational sampling, two clusters around 82.8 and 90\,days are found. The datasets producing periods with FAP$<$0.1\% correspond to the differential line width computed by \texttt{SERVAL} from VIS and NIR spectra \citep{Zechmeister2018}, and the FeH Wing-Ford band index from the NIR band \citep{Schofer2021}. This  further confirms that the rotation period of this system is $\sim$85 days.

We applied the same methodology for the other targets analysed in this paper. For GJ\,317 and GJ\,508.2, no significant periods were found in any of the activity indicators. On a cautionary note, the low number of CARMENES spectroscopic observations may make it difficult to detect rotation periods of several tens of days as expected for such stars (see Table\,\ref{tab:stars}). For GJ\,463, no cluster of periods is identified. Only two sets of indicators have FAPs below the 0.1\% level, the pseudo-equivalent width of the \ion{He}{i} $\lambda$10,833\,\AA\ line (from the NIR spectra) and the Ca\,IRT\,b line index, with peaks at periods of 56.2\,days (FAP=0.02\%) and 51.8\,days (FAP=0.05\%), in close agreement with the peaks reported in \cite{Shan2024}. The bottom panel in Fig.\,\ref{fig:clustering} shows the clustering periodogram for the activity indices of GJ\,9773. A cluster at a period $\sim$1200\,days is identified using {\tt DBSCAN}. The peaks in this cluster correspond to CaH$_{3}$, Ca\,IRT lines, and the chromatic index of the VIS spectra, the two latter indices being the only ones with a significant FAP$<$0.1\%. Such a period is far from the period of 2965\,days reported for the planet candidate (see Section\,\ref{sec:GJ9773_rv}) and the $\sim$56-day period measured from photometry (see next section). The signal present in these indicators may be attributed to an activity cycle, but the lack of similar signatures in other chromospheric indices or the RVs does not provide conclusive evidence.

\subsection{Analysis of photometric time series}
\label{sec:activity_photometry}
As reported in Section\,\ref{sec:observations}, we gathered photometric time series of GJ\,9773 and GJ\,508.2 to measure their rotation periods. Before performing the periodogram analysis, we computed nightly averages for each dataset and we applied a 3$\sigma$ clipping filter to the averages and to the uncertainties to remove obvious outliers or nights showing large photometric dispersion. The resulting photometric time series are provided in Tables\,\ref{tab:LC_GJ9773} and \ref{tab:LC_GJ508}. We subsequently computed the periodogram for each dataset separately and also a joint periodogram of those showing significant signals (FAP$<$10\%) by adjusting a different jitter term for each light curve. Table\,\ref{tab:phot} lists the number of epochs ($N$), the time span of the observations ($\Delta T$), the standard deviation ($\sigma$), the main peak found in the periodogram for each dataset, and its estimated FAP.

\begin{table}[t]
\caption[]{Periodogram analysis of the photometric time series of GJ\,9773.}
\label{tab:phot}
\begin{tabular}{rccccc}
\hline
\hline
\noalign{\smallskip}
                 &     &         & $\Delta T$ & $P_{\rm rot}$& FAP \\
Dataset          & $N$ &$\sigma$ & (days)     & (days)     & (\%) \\
\noalign{\smallskip}
\hline
\noalign{\smallskip}
\multicolumn{6}{c}{GJ\,9773}  \\
\noalign{\smallskip}
\hline                                                     
\noalign{\smallskip}
LCOGT $B$-band     & 117 & 0.0066  & 233.8      &  54.11     & 0.05    \\
LCOGT $V$-band     &  59 & 0.0037  & 207.7      &  58.87     & 5.7  \\
OSN $V$-band       &  78 & 0.0040  & 192.6      &  54.14     & 0.001   \\
EYE $V$-band       & 143 & 0.015   & 269.6      &  60.63     & 0.51    \\
OSN $R$-band       &  76 & 0.0035  & 192.6      &  107.50    & 0.001   \\
TJO $R$-band       &  21 & 0.0048  &  84.0      &  --        & --       \\
EYE $R$-band       & 139 & 0.013   & 256.6      &  --        & --       \\
\noalign{\smallskip}
\hline
\noalign{\smallskip}
\multicolumn{6}{c}{GJ\,508.2} \\
\noalign{\smallskip}
\hline 
\noalign{\smallskip}
LCOGT $B$-band     &  92 & 0.0049  & 263.6      & 37.75      & 1.3 \\
LCOGT $V$-band     &  64 & 0.0025  & 118.8      & --         & --  \\
EYE $V$-band       &  13 & 0.0050  & 69.0       & --         & -- \\
TJO $R$-band       &  87 & 0.0061  & 263.8      & --         & --  \\
EYE $R$-band       &  91 & 0.010   & 246.6      & --      & --  \\
\noalign{\smallskip}
\hline
\end{tabular}

\textbf{Notes:} Only $P_{rot}$ values with FAP$<$10\% are listed in this table.
\end{table}

\begin{figure*}[t]
\centering
\includegraphics[width=0.9\textwidth]{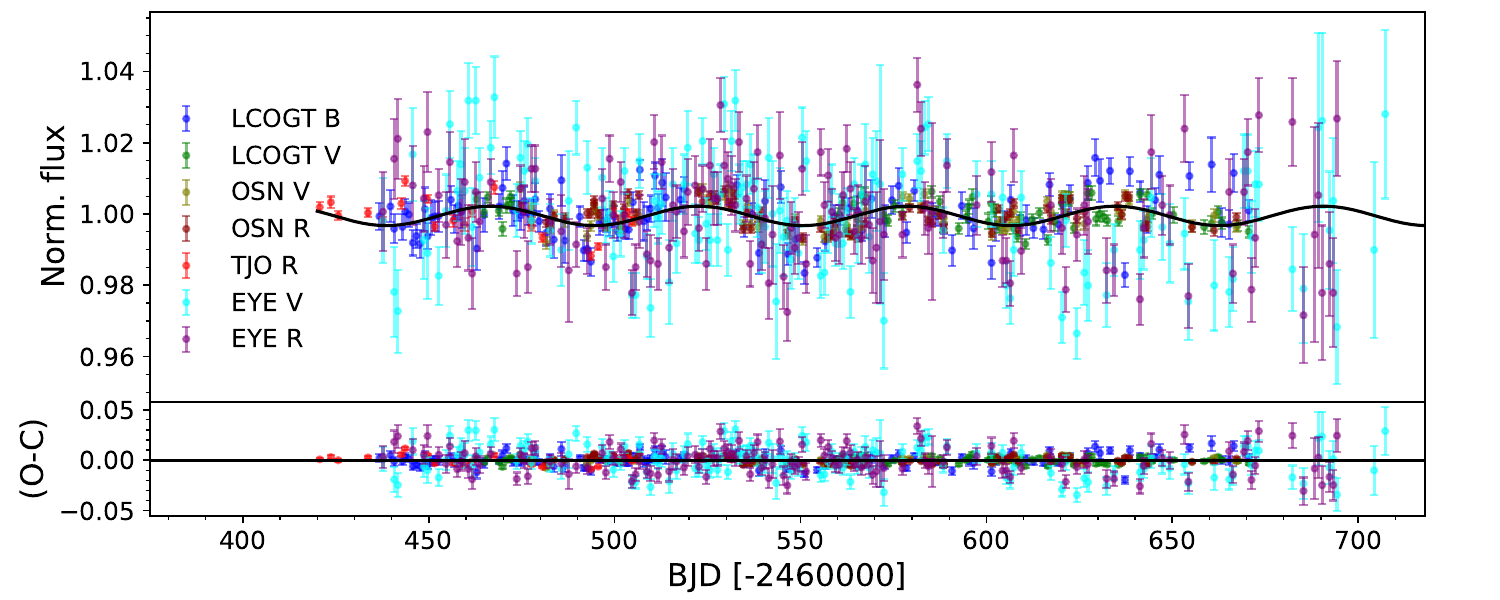}
\includegraphics[width=0.9\textwidth]{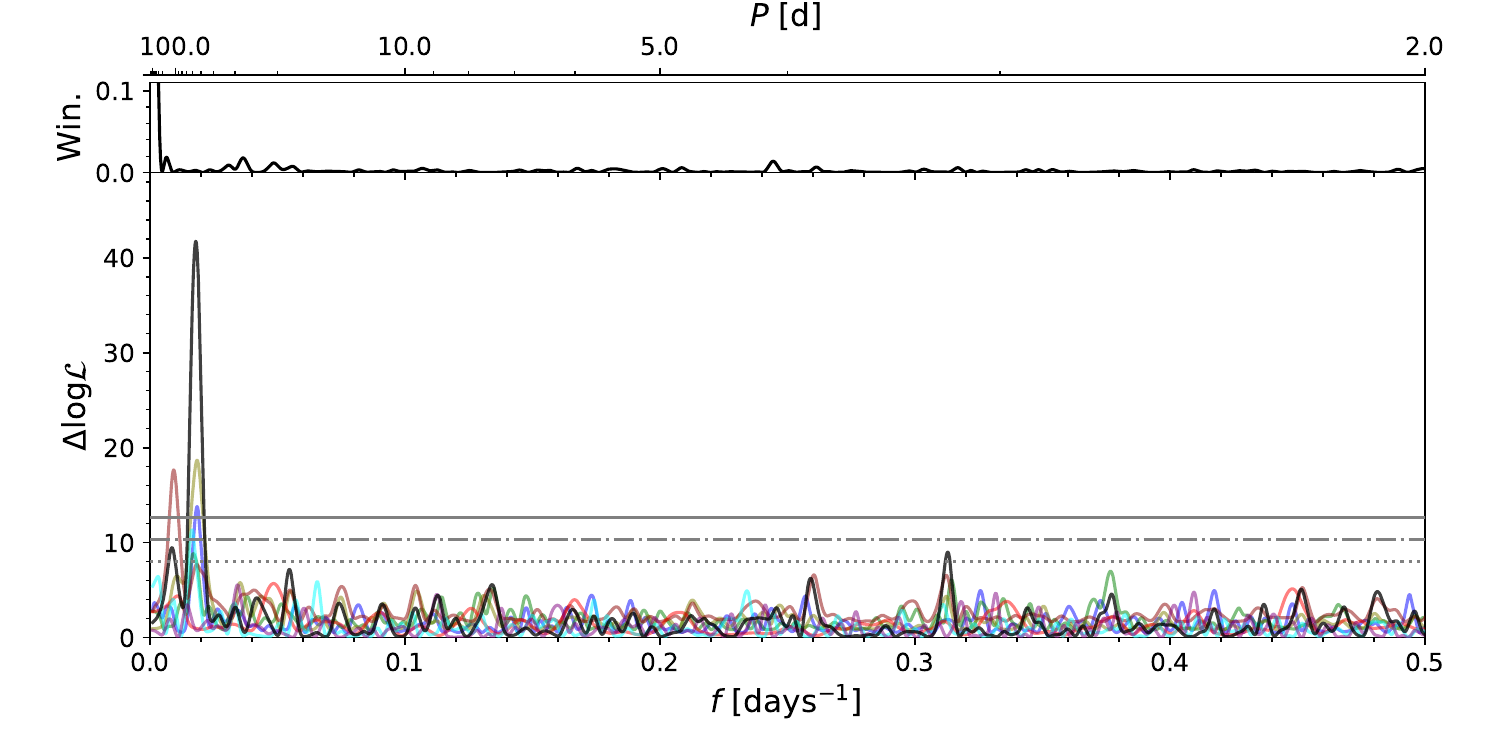}
\caption{Top: Time series of the GJ\,9773 photometry, from LCOGT, TJO, and EYE observatories. Symbols with different colours correspond to different datasets as labelled. The solid black line shows the sinusoidal model that simultaneously best fits all the datasets to the period determined from LCOGT-$B$, OSN-$V$ and $R$, and EYE-$V$ datasets. The residuals of this fit are plotted in the bottom panel. Bottom: GLS periodogram of the analysed light curves. Different colours are used to illustrate the periodogram of each dataset, while the solid black lines depict the joint periodogram for LCOGT-$B$, OSN-$V$ and $R$, and EYE-$V$ datasets. Solid, dot-dashed, and dotted horizontal grey lines indicate the 10\%, 1\%, and 0.1\% FAP levels for the joint periodogram. The corresponding window function is shown in the top panel of this figure.}
\label{fig:phot_GJ9773}
\end{figure*}

The light curves and periodogram analysis of GJ\,9773 are illustrated in the top and bottom panels of Fig.\,\ref{fig:phot_GJ9773}, respectively. The LCOGT $B$-band and OSN $V$-band periodograms display prominent peaks at a period of $\sim$54.1\,days while the EYE $V$-band and LCOGT $V$-band suggest a slightly larger period of 60.6\,days and 58.9\,days, respectively, at lower significance (FAP$\sim$0.5\% and 5.7\%). However, interestingly, the highest peak of the OSN $R$-band periodogram is consistent with 2 times the period of the OSN $V$-band dataset, and a second peak at $\sim$54\,days is also present. The analysis of the other datasets does not result in any significant period detection. On the other hand, the highest peak of the EYE $R$-band periodogram ($P_{\rm rot}\sim$29\,days) corresponds to half the period found in EYE $V$-band dataset. The simultaneous periodogram analysis of the LCOGT-$B$ and $V$, OSN-$V$ and $R$, and EYE-$V$ datasets results in a significant broad peak at $\sim$55.7\,days with a flux semi-amplitude of $\sim$0.3\%. We adopt the central value of this broad peak and half the width at its full-width-half-maximum as the rotation period and its uncertainty for GJ\,9773, 56$\pm$8\,days. Although uncertain, this is consistent with a rather inactive old M-dwarf star \citep{Noyes1984,Barnes2003,CortesContreras2024}.

\begin{figure*}[t]
\centering
\includegraphics[width=0.9\textwidth]{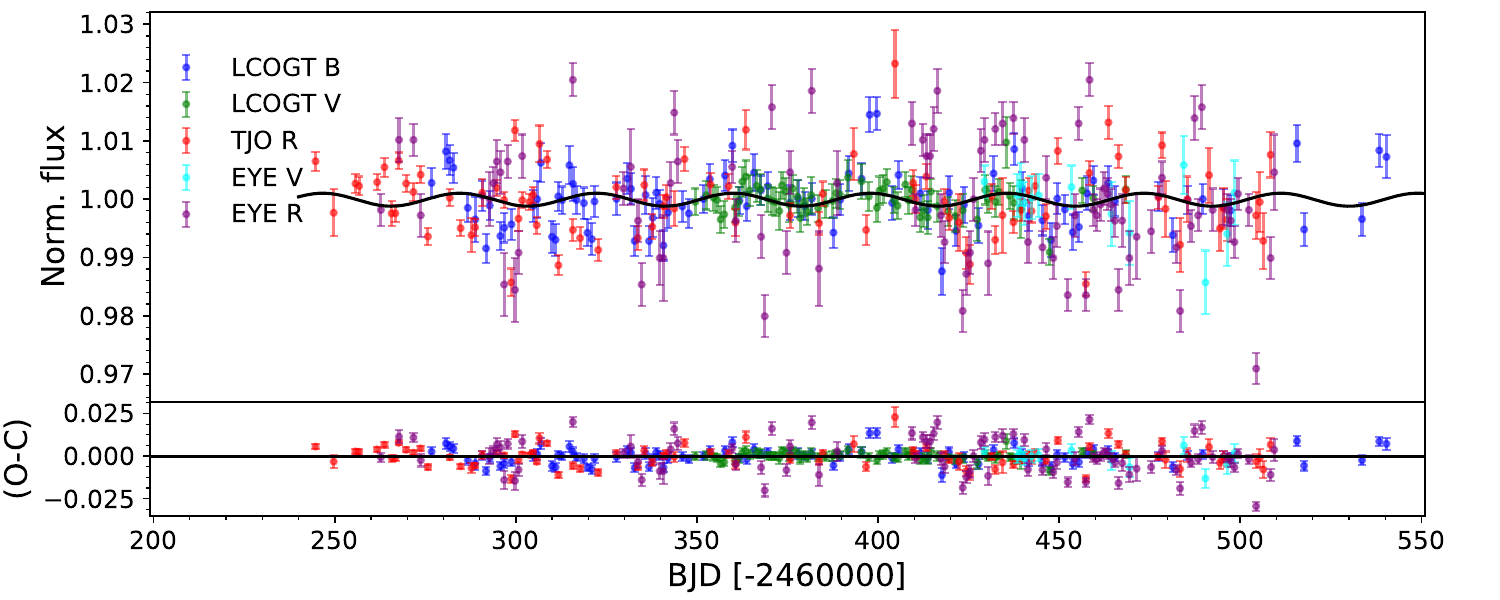}
\includegraphics[width=0.9\textwidth]{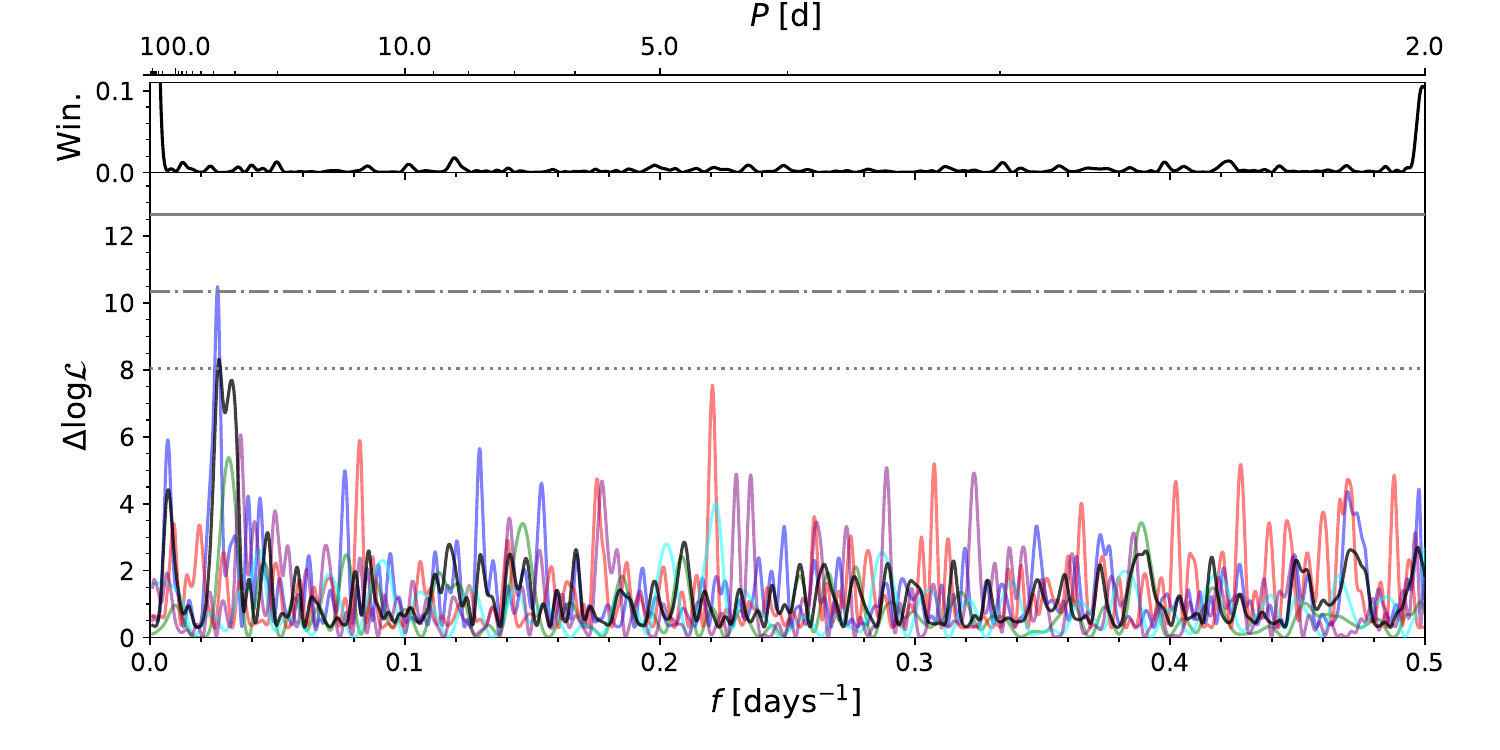}
\caption{Same as Fig.\,\ref{fig:phot_GJ9773}, but for GJ\,508.2.}
\label{fig:phot_GJ508}
\end{figure*}

Figure\,\ref{fig:phot_GJ508} shows the analysis of the GJ\,508.2 photometry. In this case, a signal with FAP$\sim$1\% is found in the periodogram of the LCO $B$-band dataset at $\sim$37.5\,days. The LCOGT $V$-band data shows a clear peak at a shorter period of $\sim$32.4\,days, although with a significance below the 10\% FAP level. The periodograms of all the other datasets are less clear, showing several peaks with similar significance from a few to several tens of days. The highest peak of the joint LCOGT-$B$ and LCOGT-$V$ periodogram corresponds to a period of 36.9 days with a FAP of only $\sim$9\% and double structure. Although we cannot draw firm conclusions from this analysis, following the same approach as for the target discussed before, we adopt 36$\pm$6\,days as the most probable rotation period for GJ\,508.2. This value is slightly shorter but consistent within the uncertainties with the period reported by \cite{Oelkers2018}, and also points towards a rather inactive old M-dwarf star. However, we note here that this value remains to be confirmed with further observations.

\section{Radial velocity analysis}
\label{sec:RVanalysis}
The RV time series were analysed using \texttt{Exo-Striker} \citep{exostriker} to infer initial values of the orbital properties of the planets and also with the \texttt{Juliet} package \citep{Espinoza2019} making use of the RV fitting \citep{Fulton2018}, and nested sampling \citep{Feroz2009} functionalities to compute the posterior distributions and estimate uncertainties. For each system, we calculated a GLS periodogram \citep{Zechmeister2009} to obtain a first approximation of the variability period and to search for additional significant signals in the data after pre-whitening. The periods of all reported planetary companions were recovered at this step. Besides, we inspected the residuals of the fits for additional signals. The evidence ($\mathcal{Z}$) and likelihood ($\mathcal{L}$) parameters were used to compare different solutions and to assess the significance of additional Keplerian components in the data. We also added linear and quadratic trends to our RV models to account for possible very-long period signal, but such solutions were not preferred for any of the systems analysed here based on $\mathcal{Z}$.

For each planet candidate, we fitted the orbital period ($P$), the time of conjunction (BJD$_{0}$), the RV semi-amplitude ($K$), and the eccentricity ($e$) and argument of periastron ($\omega$) parameterised as $e \cos \omega$ and $e \sin \omega$. The RV values from different instruments were fitted simultaneously for each target, and thus RV offsets ($\gamma$) and jitter terms ($\sigma$) were considered as free parameters for each instrument time series. Wide uniform priors were used for all parameters, except for the jitter, for which a log-normal distribution was employed. In the case of multi-planetary systems, non overlapping period ranges were set as priors for each planet once identified from the periodogram and the preliminary \texttt{Exo-Striker} analysis. The orbital parameters obtained as the median and 1$\sigma$ percentiles of the posterior distributions obtained using \texttt{Juliet} are reported in Table\,\ref{tab:RVfit}. The derived physical quantities for each planetary system are also listed in this table. A detailed description for each target is provided below.

\begin{table*}[t]
\centering
\caption[]{Adopted best fit parameters and derived exoplanet and orbital properties of the systems analysed here.}
\label{tab:RVfit}
\begin{tabular}{rccccc}
\hline
\hline
\noalign{\smallskip}
Parameter                & GJ 3512                           & GJ 317                      & GJ 463                      & GJ 9773                     & GJ 508.2                     \\
\noalign{\smallskip}
\hline                                                       \noalign{\smallskip}
\multicolumn{6}{c}{Fitted parameters}\\                      
\noalign{\smallskip}
\hline
\noalign{\smallskip}
$P_{\rm b}$ (d)              & 203.109\aunc{0.035}{-0.034}       & 695.69\aunc{0.26}{-0.25}    & 3459$\pm$22                 & 2965\aunc{41}{-37}          & >5300                        \\[0.205ex]
BJD$_{0,\rm b}$ (+2450000)   & 7739.51\aunc{0.22}{-0.20}         & 8089.4\aunc{1.7}{-1.6}      & 4313\aunc{28}{-30}          & 9362$\pm$18                 & --                           \\[0.205ex]
$K_{\rm b}$ (\ms)            & 71.28\aunc{0.32}{-0.31}           & 71.39\aunc{0.56}{-0.55}     & 36.27$\pm$0.71              & 15.11\aunc{0.32}{-0.31}     & >33                          \\[0.205ex]
$e_{\rm b} \cos \omega_{b}$  &--0.2563\aunc{0.0041}{-0.0040}     & 0.0808\aunc{0.0067}{-0.0070}& 0.125\aunc{0.013}{-0.014}   & 0.034\aunc{0.020}{-0.019}   & --                           \\[0.205ex]
$e_{\rm b} \sin \omega_{b}$  & 0.3427\aunc{0.0041}{-0.0040}      & 0.0159\aunc{0.0071}{-0.0069}&--0.074$\pm$0.016            & 0.214$\pm$0.021             & --                           \\[0.205ex]
$P_{\rm c}$ (d)              & 2354\aunc{36}{-30}                & 6500\aunc{169}{-166}        & --                          & --                          & --                           \\[0.205ex]
BJD$_{0,\rm c}$ (+2450000)& 7339\aunc{32}{-37}               & 12735\aunc{179}{-176}       & --                          & --                          & --                           \\[0.205ex]
$K_{\rm c}$ (m\,s$^{-1}$)    & 28.02\aunc{0.36}{-0.34}           & 29.33\aunc{0.74}{-0.73}     & --                          & --                          & --                           \\[0.205ex]
$e_{\rm c} \cos \omega_{\rm c}$  & 0.056\aunc{0.016}{-0.015}         &--0.062\aunc{0.020}{-0.021}  & --                          & --                          & --                           \\[0.205ex]
$e_{\rm c} \sin \omega_{\rm c}$  & 0.067\aunc{0.012}{0.013}          & 0.207$\pm$0.013             & --                          & --                          & --                           \\[0.205ex]
$\mu_{\rm VIS}$ (\ms)    & 11.41\aunc{0.40}{-0.42}           & 6.9$\pm$1.6                 & 9.42\aunc{0.48}{-0.46}      &--4.93$\pm$0.20              & --                           \\[0.205ex]
$\mu_{\rm NIR}$ (\ms)    & 11.60\aunc{0.67}{-0.65}           & 12.8$\pm$1.9                & 11.30\aunc{0.71}{-0.75}     &--5.35$\pm$0.61              & --                           \\[0.205ex]
$\gamma_{\rm IRD}$ (\ms)    & --16.6\aunc{1.1}{-1.2}            & --                          & --                          & --                          & --                        \\[0.205ex]
$\gamma_{\rm HIRES}$ (\ms)  & --                                &--12.7$\pm$1.0               & 10.8$\pm$2.2                & --                          & --                        \\[0.205ex]
$\gamma_{\rm PFS}$ (\ms)    & --                                & 28.2$\pm$1.1                &                             & --                          & --                        \\[0.205ex]
$\gamma_{\rm HARPS-pre}$ (\ms)& --                              &--19.53\aunc{0.94}{-0.97}    & --                          & --                          & --                        \\[0.205ex]
$\gamma_{\rm HARPS-post}$ (\ms)& --                             & 41.3$\pm$1.6                & --                          & --                          & --                        \\[0.205ex]
$\gamma_{\rm HRS}$ (\ms)    & --                                & --                          & 12.8$\pm$1.1                & --                          & --                        \\[0.205ex]
$\sigma_{\rm VIS}$ (\ms) & 1.95\aunc{0.23}{-0.24}            & 2.05\aunc{0.44}{-0.38}      & 0.02\aunc{0.19}{-0.02}      & 0.17\aunc{0.60}{-0.16}      & --                           \\[0.205ex]
$\sigma_{\rm NIR}$ (\ms) & 4.86\aunc{0.70}{-0.72}            & 4.3\aunc{1.6}{-2.2}         & 4.39\aunc{0.75}{-0.78}      & 3.1\aunc{0.8}{-1.1}         & --                           \\[0.205ex]
$\sigma_{\rm IRD}$ (\ms) & 5.4\aunc{1.0}{-0.9}               & --                          & --                          & --                          & --                           \\[0.205ex]
$\sigma_{\rm HIRES}$ (\ms)& --                               & 4.41\aunc{0.68}{-0.62}      & 7.6\aunc{2.1}{-1.6}         & --                          & --                           \\[0.205ex]
$\sigma_{\rm PFS}$ (\ms) & --                                & 3.63\aunc{0.89}{-0.76}      & --                          & --                          & --                           \\[0.205ex]
$\sigma_{\rm HARPS-pre}$ (\ms)& --                           & 2.86\aunc{0.33}{-0.31}      & --                          & --                          & --                           \\[0.205ex]
$\sigma_{\rm HARPS-post}$ (\ms)&                             & 4.25\aunc{0.59}{-0.50}      & --                          & --                          & --                           \\[0.205ex]
$\sigma_{\rm HRS}$ (\ms) & --                                & ---                         & 0.05\aunc{0.71}{-0.05}      & --                          & --                           \\[0.20ex]
\noalign{\smallskip}
\hline
\noalign{\smallskip}
\multicolumn{6}{c}{Residuals}\\
\noalign{\smallskip}
\hline
\noalign{\smallskip}
rms$_{\rm VIS}$ (\ms)    &  3.48                             &  3.27                       &  3.18                       &  2.21                       & --                           \\[0.0ex]
rms$_{\rm NIR}$ (\ms)    & 10.99                             &  8.49                       &  8.93                       &  7.84                       & --                           \\[0.0ex]
rms$_{\rm IRD}$ (\ms)    &  6.79                             &  --                         &  --                         &  --                         & --                           \\[0.0ex]
rms$_{\rm HIRES}$ (\ms)  &  --                               &  5.54                       &  8.34                       &  --                         & --                           \\[0.0ex]
rms$_{\rm PFS}$ (\ms)    & --                                &  4.88                       &  --                         &  --                         & --                           \\[0.0ex]
rms$_{\rm HARPS-pre}$ (\ms) & --                             &  3.44                       &  --                         &  --                         & --                           \\[0.0ex]
rms$_{\rm HARPS-post}$ (\ms) & --                            &  4.69                       &  --                         &  --                         & --                           \\[0.0ex]
rms$_{\rm HRS}$ (\ms)    & --                                &  --                         &  6.82                       &  --                         & --                           \\[0.0ex]
\noalign{\smallskip}
\hline
\noalign{\smallskip}
\multicolumn{6}{c}{Computed properties}\\
\noalign{\smallskip}
\hline
\noalign{\smallskip}
$e_{b}$                       & 0.4279\aunc{0.0036}{-0.0035} &0.0826\aunc{0.0070}{-0.0071} &0.146\aunc{0.014}{-0.015}    &0.218$\pm$0.020              & >0.27                        \\[0.205ex]
$\omega_{\rm b}$ (deg)            & 126.80\aunc{0.62}{-0.64}     &11.3\aunc{4.8}{-4.6}         &329.3\aunc{5.9}{-5.8}        &99.0$\pm$5.2                 & >302                         \\[0.205ex]
$a_{2,\rm b}$ (au)                & 0.3376\aunc{0.0082}{-0.0084} &1.138$\pm$0.013              &3.530\aunc{0.037}{-0.038}    &2.921$\pm$0.045              & >4.85                        \\[0.205ex]
$m_{\rm b} \sin i_{b}$ (M$_{\rm Jup}$) & 0.461$\pm$0.023   &1.688$\pm$0.041              &1.649\aunc{0.044}{-0.045}    &0.543$\pm$0.016              & >1.85                        \\[0.205ex]
$e_{\rm c}$                       & 0.089$\pm$0.012              &0.217$\pm$0.014              & --                          & --                          & --                           \\[0.205ex]
$\omega_{\rm c}$ (deg)            & 50$\pm$10                    &106.7\aunc{5.4}{-5.3}        & --                          & --                          & --                           \\[0.205ex]
$a_{2,\rm c}$ (au)                & 1.731\aunc{0.043}{-0.046}    &5.04$\pm$0.10                & --                          & --                          & --                           \\[0.205ex]
$m_{\rm c} \sin i_{c}$ (M$_{\rm Jup}$) & 0.453$\pm$0.023   &1.430\aunc{0.058}{-0.055}    & --                          & --                          & --                           \\[0.205ex]
\noalign{\smallskip}
\hline
\noalign{\smallskip}
\end{tabular}
\end{table*}

\subsection{GJ\,3512}
\label{sec:GJ3512_rv}
\begin{figure*}[t]
\centering
\includegraphics[width=1.0\textwidth]{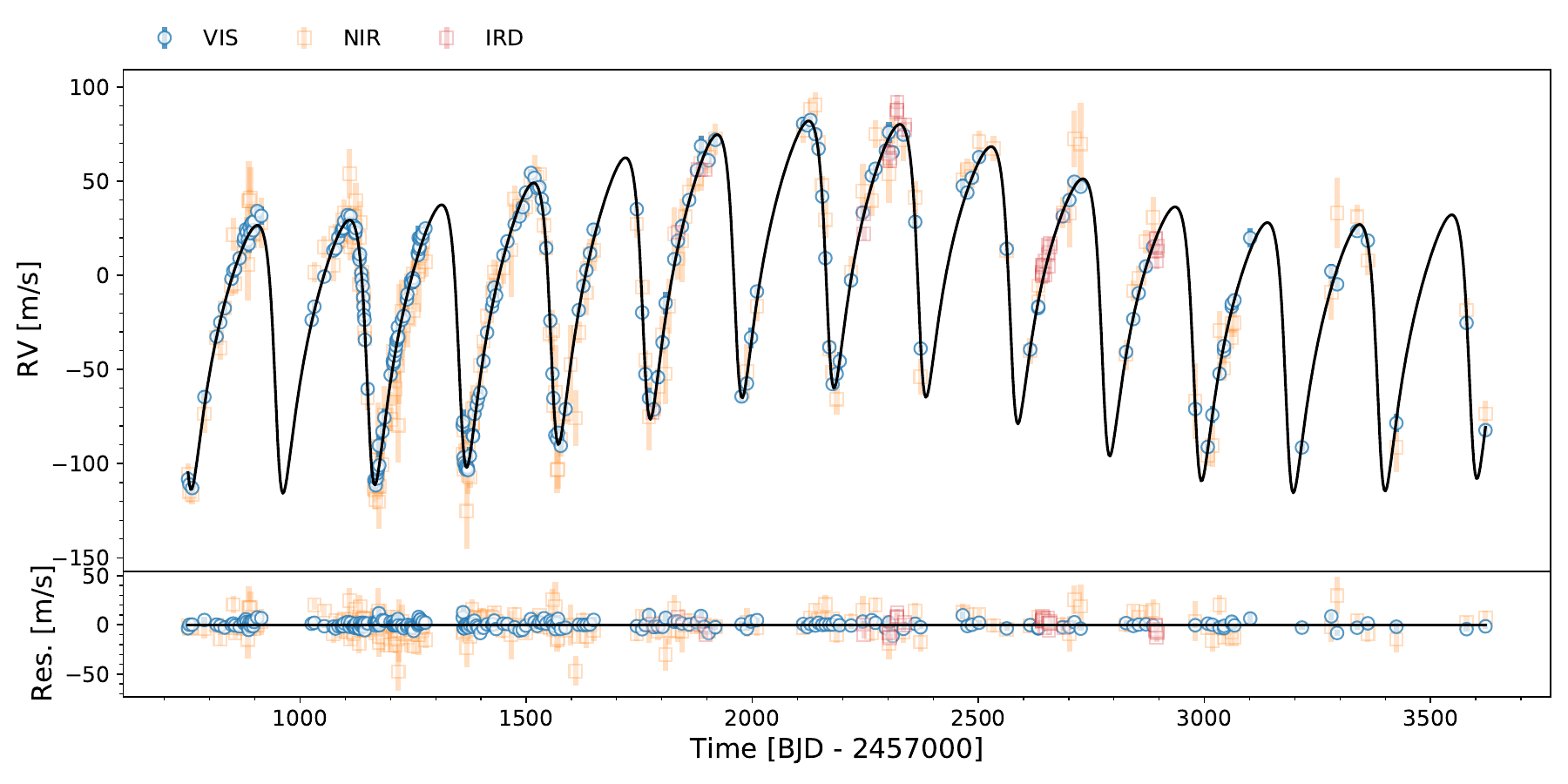}
\includegraphics[width=0.43\textwidth]{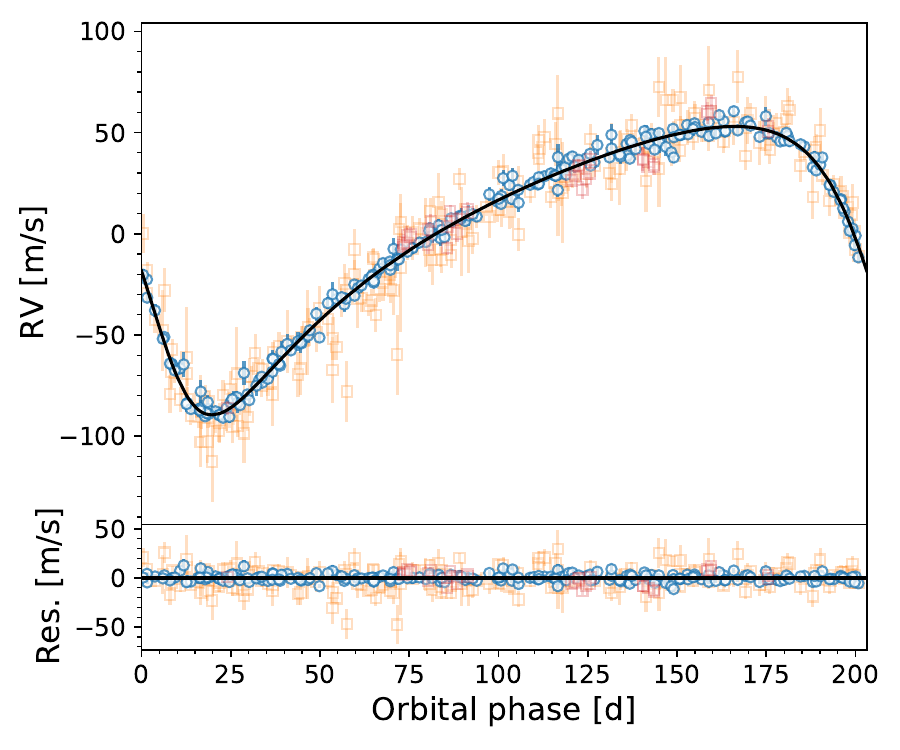}
\includegraphics[width=0.43\textwidth]{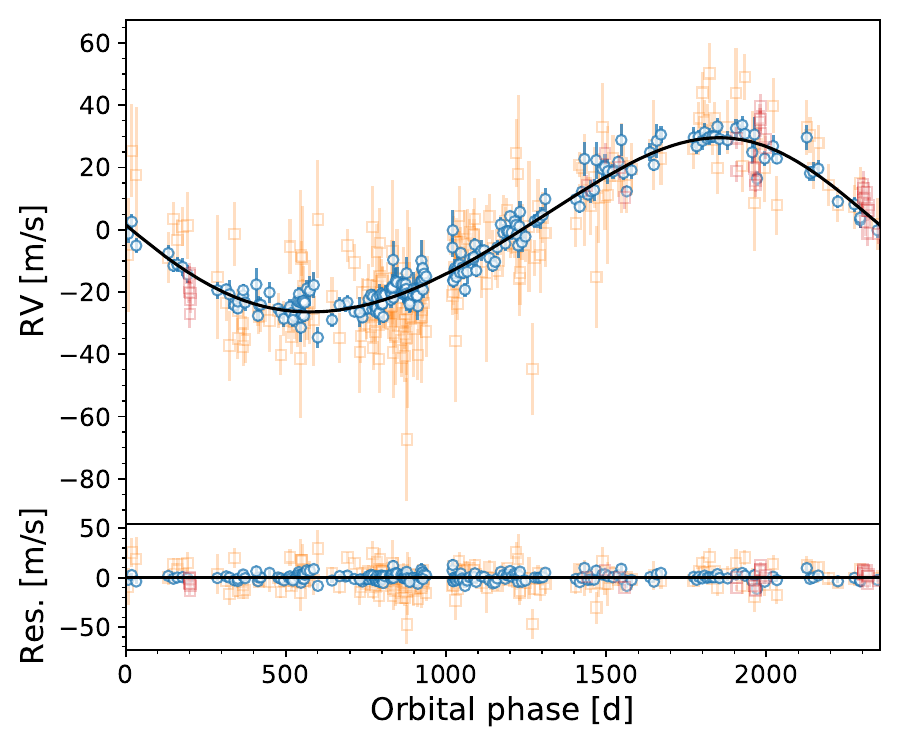}
\caption{Top: RV curve of GJ\,3512 with the best Keplerian fit shown as a solid black line. RV measurements from the CARMENES VIS and NIR channels are shown as blue circles and orange squares, respectively, while data from IRD are illustrated as red squares. Residuals are shown in the bottom panel of this plot. Bottom: Phase-folded radial velocity curve of GJ\,3512 caused by planet b (left) and c (right).}
\label{fig:rv_gj3512}
\end{figure*}

The CARMENES (VIS and NIR) and IRD RVs time series of GJ\,3512 are consistent with two Keplerian orbits with periods of $\sim$203.1 and $\sim$2354 days and RV semi-amplitudes of 71.28 and 28.02\,\ms, respectively. The top panel in Figure\,\ref{fig:rv_gj3512} shows the best fit to the RV time series. The bottom panels illustrate the phase-folded RV variations caused by each planet independently. Compared with the previous study of \cite{Morales2019}, the new CARMENES and IRD data extend the RV monitoring of this system by $\sim$2000 days, closing a full cycle for the outer planet candidate, thus firmly constraining its orbital period. Circular and eccentric models for the outer planet were tested, with the slightly eccentric solution ($e \sim 0.09$) being significantly better ($\Delta \ln \mathcal{Z} = 20$). 

\begin{figure}[t]
\centering
\includegraphics[width=\columnwidth]{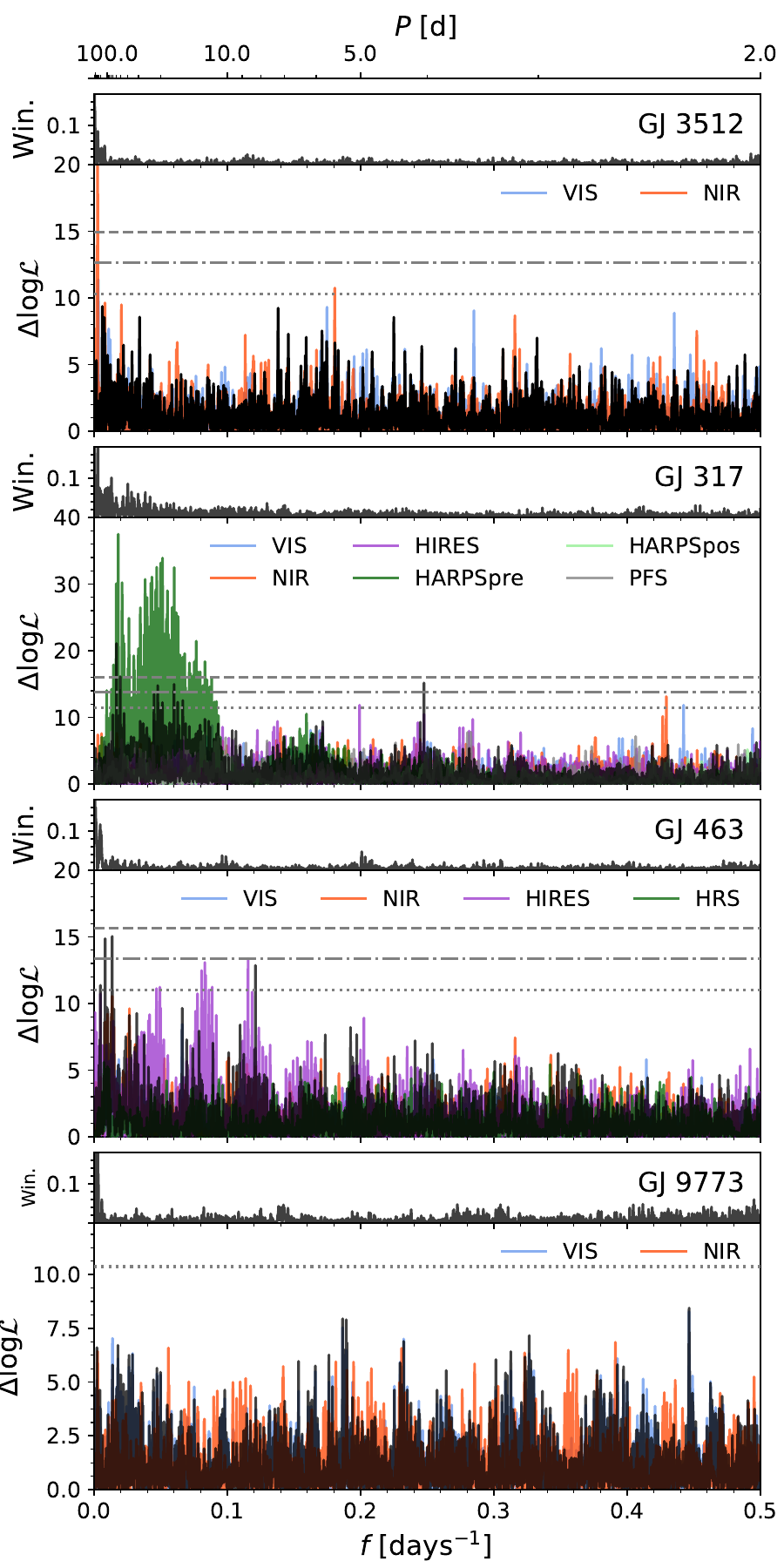}
\caption{Periodogram of the residuals from the fits to the RV datasets of GJ\,3512, GJ\,317, GJ\,463, and GJ\,9773 (from top to bottom). In each panel the solid black line corresponds to the joint periodogram of all datasets, while solid lines depict the periodogram for each individual dataset as labelled. The top panel for each target illustrates the window function. Solid horizontal grey lines show the 0.1\% (dashed), 1\% (dot-dashed), and 10\% (dotted) FAP levels.}
\label{fig:res_periodogram}
\end{figure}

In order to check for further signals, we computed the residuals of the best fit to each dataset, and we performed a joint periodogram analysis. The top panel of Fig.\,\ref{fig:res_periodogram} shows that no further significant signals are found in the residuals with a false alarm probability (FAP) below 0.1\%. The most prominent peaks in the joint periodogram correspond to periods of $\sim$7.2 and $\sim$130 days, which may be related to aliases of the rotation period and the sampling. As a further test, we added to the fits a Gaussian process  component (GP) to take into account any possible intrinsic variability caused by stellar rotation.  We tested double simple harmonic oscillator \citep[dSHO,][]{ForemanMackey2017} and quasi-periodic with cosine \citep[QPC,][]{Perger2021} kernels. We set a normal prior to the rotation period of the star (85$\pm$15 days). The parameters of the fit were not substantially changed with respect to those reported in Table\,\ref{tab:RVfit}, and although at a different period, the GP components reduce the significance of the signals at $\sim$7.2 and $\sim$130 days.

The inner planet of this system was announced in \cite{Morales2019}. We can now improve the accuracy of its parameters thanks to the longer time span of the data and the more precise determination of the orbit of the outer system. The period of the orbit of the inner planet that we report here is $\sim$0.5\,days shorter, but the parameters that we obtain are consistent within 2$\sigma$ uncertainties. In accordance with the dynamical analysis in \cite{Morales2019}, the nearly circular orbit of the outer planet agrees well with the predictions. In conclusion, GJ\,3512 turns out to be a multi-planet system with two very similar gas giants of $\sim$0.45\,\mjup. The rms of the RV residuals, in particular that of the more precise CARMENES VIS data (3.48 \ms), indicate that no further planets with masses greater than that of Saturn are present in the system at periods shorter than $\sim$40 years. On the other hand, injection and retrieval tests on the residuals of the VIS RVs indicate that there are no further planets inside the orbit of GJ\,3512\,b with a mass above $\sim$3\,M$_{\oplus}$ (see Fig.\,\ref{fig:detection_limit_GJ3512}).

\subsection{GJ\,317}
\label{sec:GJ317_rv}
\begin{figure*}[t]
\centering
\includegraphics[width=1.0\textwidth]{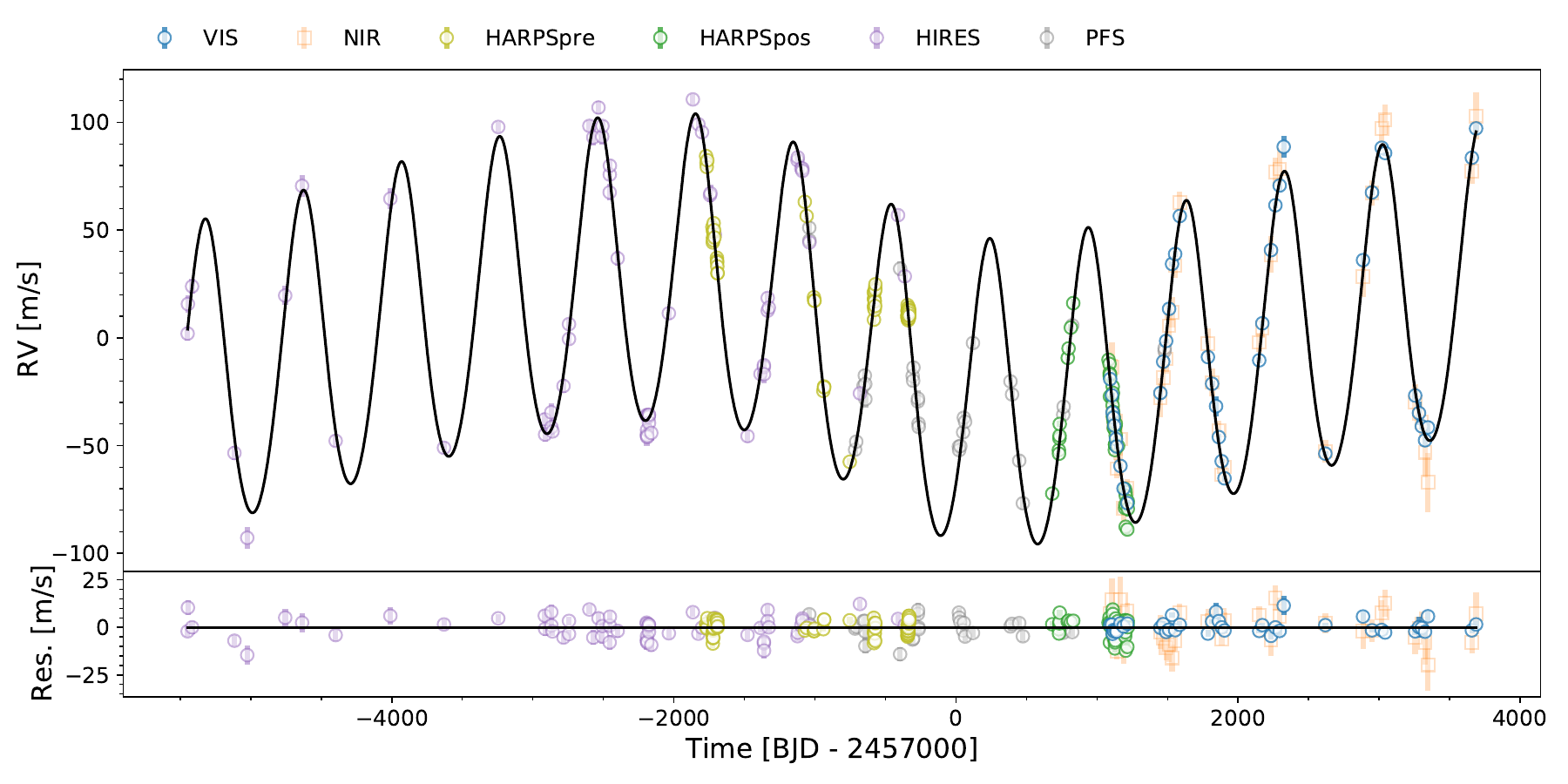}
\includegraphics[width=0.43\textwidth]{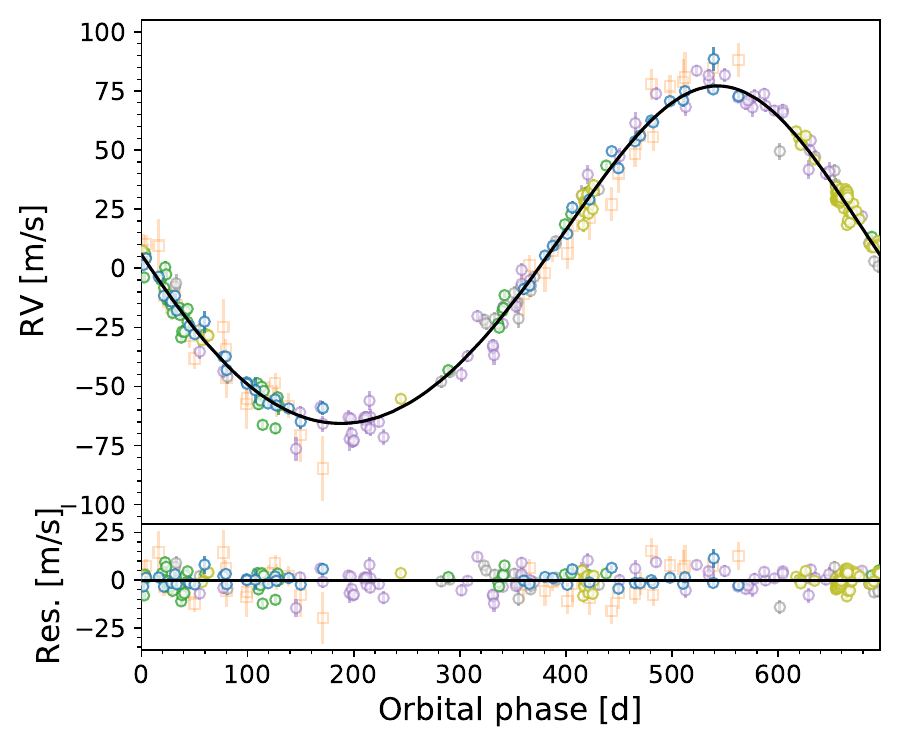}
\includegraphics[width=0.43\textwidth]{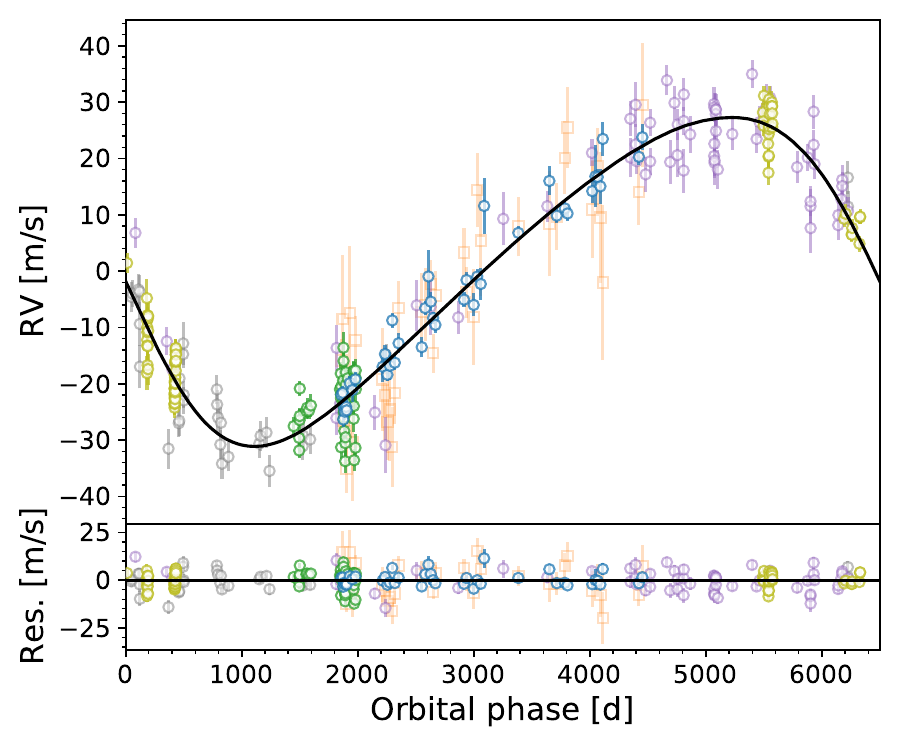}
\caption{Top: RV curve of GJ\,317 with the best Keplerian orbit fit shown as a solid black line. RV measurements from the CARMENES VIS and NIR channels are depicted as blue circles and orange squares, respectively. Data from HIRES, HARPS (pre and pos), and PFS are illustrated with different colours as labelled. Residuals are shown in the panel underneath. Bottom: Phase-folded RV curves of GJ\,317\,b (left) and c (right).}
\label{fig:rv_gj317}
\end{figure*}

For GJ 317, we fitted simultaneously the RVs from CARMENES (VIS and NIR), HARPS, HIRES, and PFS with a two planet model based on the orbital parameters of \cite{Feng2020}. HARPS data were split in two parts to account for possible offsets caused by the instrument intervention in December 2016 \citep[see][]{LoCurto2015}. The best fit results in an offset of $\sim$60.83\,\ms\ between data before (HARPSpre) and after (HARPpos) this intervention. Figure\,\ref{fig:rv_gj317} depicts the best fit to the RV time series of this planetary system (top panel) and the phased-folded RV curve for each planet independently (bottom panels). CARMENES RVs extend the time series baseline by almost 1900 days, sampling partially a second orbital cycle of the outer planet. The orbital periods determined for the inner and outer planets are 695.6 and 6500 days, respectively, which are consistent within uncertainties with those reported by \cite{AngladaEscude2012} and \cite{Feng2020}, although we obtain a period for GJ\,317\,c that is shorter by $\sim$240\,days. This period difference is in agreement with the wide range for the period of planet c reported by \cite{AngladaEscude2012} (6100 to 15100 days) and within 1$\sigma$ of the combined uncertainties of the $6739\pm143$\,days listed in \cite{Feng2020}.

The periodogram of the residuals of the two planet fit, illustrated in the second panel in Fig.\,\ref{fig:res_periodogram}, still shows significant power at a period of 59.5\,days (FAP<0.1\%) and a cluster of peaks around 20 days, which are dominated by the residuals from of the CARMENES VIS and HARPSpre datasets. The first value is close to the rotation period of the target (see Table\,\ref{tab:stars}), while the collection of peaks covers the harmonics at half and one third of the rotation period, which may appear depending on the spot configuration \citep[see e.g.][]{Perger2021}. We adjusted the data adding a sinusoidal function assuming a rotational period between 40 and 70 days, with a uniform prior distribution. This resulted in a significantly better fit to the RV time series ($\Delta \ln \mathcal{Z} \sim 11$) with a period of 59.42$\pm$0.06 days, and a semi-amplitude of 2.5$\pm$0.3\,\ms. We also added a GP component with a rotation period fixed to this value with a normal uncertainty of 1.5 days (see Table\,\ref{tab:stars}). However, the orbital parameters do not significantly differ from the fit that only considers two Keplerian functions that we report in the second column in Table\,\ref{tab:RVfit}.

The orbital parameters of the inner planet listed in Table\,\ref{tab:RVfit} are consistent within the uncertainties with those reported by \cite{Feng2020}. On the other hand, as stated, the analysis yields a shorter orbital period, as well as, a lower RV semi-amplitude by 2$\sigma$ for the outer planet. We also obtained a larger value and more precise measurement of the eccentricity, although still within 2$\sigma$ of the literature value. We reduced the orbital period and RV semi-amplitude uncertainties by a factor of 2 and 3, respectively, with respect to the values in \cite{Feng2020}. From the fit results, we estimate the minimum mass of the planets to be 1.688\,\mjup\, and 1.430\,\mjup\, for the inner and outer planets, respectively. These values are $\sim$3.7\% and $\sim$16\% smaller than those reported by \cite{Feng2020}. In addition to the differences in the orbital parameters that we determine for the outer planet, the mass difference is also partially due to the lower stellar mass (by $\sim$4.5\%) that we derive from our spectroscopic analysis of the CARMENES data following \cite{Schweitzer2019} with respect to that in \cite{Feng2020}.

\subsection{GJ\,463}
\label{sec:GJ463_rv}

\begin{figure*}[t]
\centering
\includegraphics[width=\textwidth]{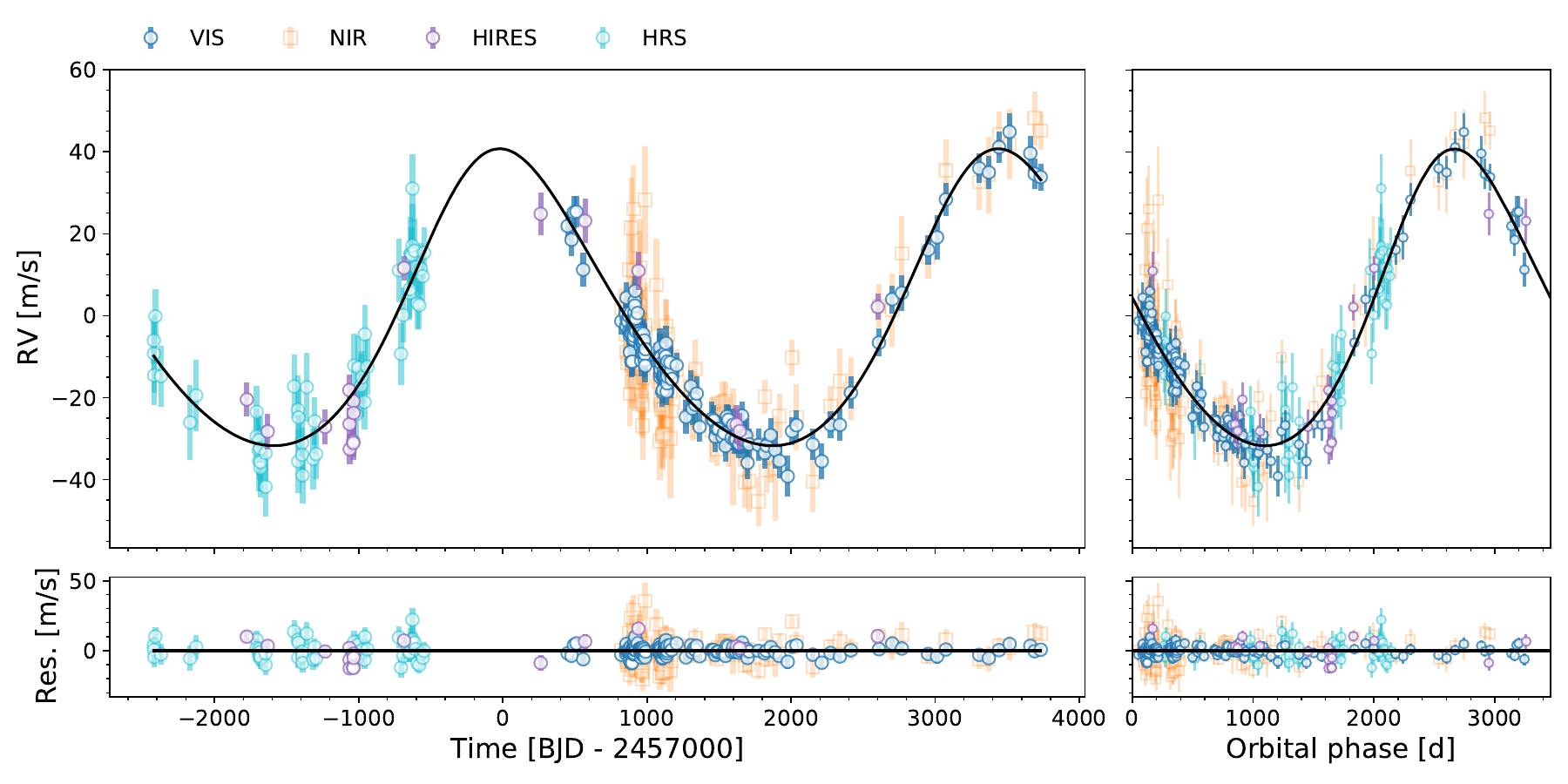}
\caption{Left: RV time series of GJ\,463 with the best Keplerian fit shown as solid black line. RV measurements from the CARMENES VIS and NIR channels are depicted as blue circles and orange squares, respectively, while data from HRS and HIRES are illustrated as cyan circles and purple squares, respectively. Residuals are shown in the bottom panel. Left: Phase-folded RV curve of GJ\,463.}
\label{fig:rv_gj463}
\end{figure*}

Following the same procedures as for the targets above, we fitted the CARMENES (VIS and NIR channels), HIRES, and HRS GJ\,463 RV time series simultaneously.  Figure\,\ref{fig:rv_gj463} illustrates the best fit model to the RV measurements (left panel) and the phase-folded RV curve (righ panel). The best-fitting parameters are listed in the third column of Table\,\ref{tab:RVfit}. Interestingly, HIRES data are contemporaneous to both HRS and CARMENES data, which  us to determine the offsets between the different datasets well. Besides, CARMENES data nearly cover a full second orbital cycle of the planet announced in \cite{Endl2022}. The RV semi-amplitudes obtained for each instrument, including the VIS and NIR CARMENES channels, are in good mutual agreement. This fact, combined with the stability of the signal, confirms that the RV variability is most certainly caused by a planet orbiting the star and not due to a stellar activity cycle.

An additional tentative signal with a period of 11.2\,days, mainly arising from HRS data, is suggested in \cite{Endl2022}. We calculated the GLS periodogram of our datasets after removing the long-period Keplerian orbit in search for additional signals. The resulting periodogram is shown in Figure\,\ref{fig:res_periodogram}. We did not find evidence for an 11.2-day periodicity, but two peaks at 74.0 and 8.2\,days are just above FAP levels of 1\% and 10\%, respectively. If we do not consider the CARMENES NIR data, the significance of both these peaks is substantially reduced reaching a FAP above the 10\% threshold. The most prominent peak in the CARMENES VIS data is at 15.1-days, with a FAP$>$10\%. As a check, we added a sinusoidal variability to our model and tested for additional periodic signals in the data with two sets of priors: between 10 and 20 days, and 60 and 80 days. We performed a joint fit to all datasets. None of these tests resulted in a significant improvement of the fit statistic ($\Delta \ln \mathcal{Z} < 0$), either with or without the NIR data. The fact that the 15.1-day peak is close to the first harmonic of the rotation period of the star (see Table\,\ref{tab:stars}) may be indicative of it being of stellar origin and not of planetary nature.

The parameters that we obtain for this planetary system are listed in Table\,\ref{tab:RVfit}. The better precision and longer time baseline of our CARMENES data made it possible to decrease the uncertainty of the orbital period and the RV semi-amplitude by a factor of $\sim 4$. This results in a more precise minimum mass of the system of 1.649\aunc{0.044}{-0.045}\,\mjup. Our results indicate a somewhat higher orbital eccentricity as well as a $\sim$8\% larger RV semi-amplitude with respect to the values reported in \cite{Endl2022}, but in both cases still within the quoted error bars. 

\subsection{GJ\,9773}
\label{sec:GJ9773_rv}
\begin{figure*}[t]
\centering
\includegraphics[width=1.0\textwidth]{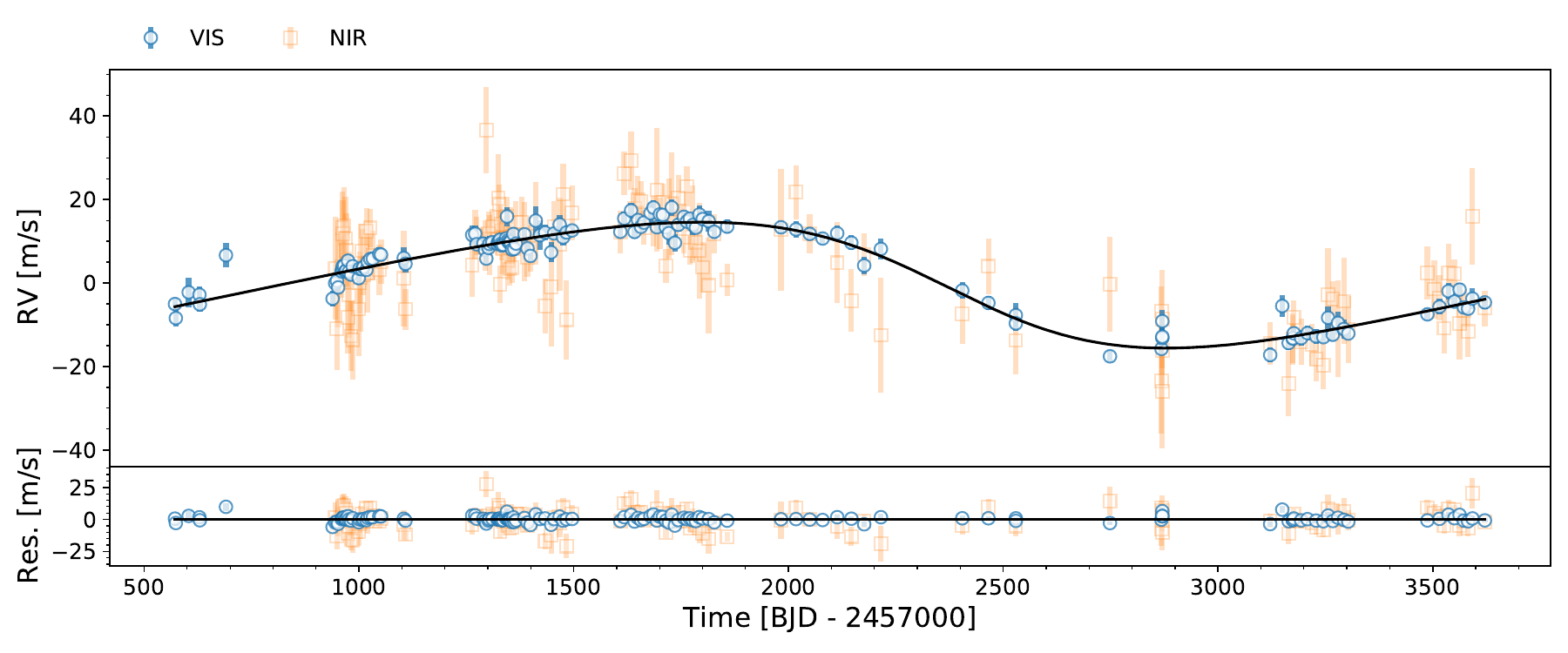}
\caption{RV curve of GJ\,9773 with the best Keplerian fit shown as solid black line. RV measurements from the CARMENES VIS and NIR channels are depicted as blue squares and orange circles, respectively. Residuals are shown in the bottom panel.}
\label{fig:rv_gj9773}
\end{figure*}

The CARMENES VIS and NIR RV time series of GJ\,9773 show variability consistent with a Keplerian orbit with a period close to the full timespan of the data. As for the previous cases, we modelled the data using \texttt{Juliet}, which yields an eccentric orbital fit with a period of $\sim$2965 days and a corresponding companion with a minimum mass of 0.543\,\mjup. Figure\,\ref{fig:rv_gj9773} illustrates the model to the RV time series from the CARMENES VIS and NIR channel spectra. The best-fitting parameters are reported in the fourth column in Table\,\ref{tab:RVfit}. No further significant periodic signals are found in the residuals as shown in the bottom panel of Fig.\,\ref{fig:res_periodogram}. Additionally, we did not find any evidence of variability at the estimated rotation period of $\sim$60\,days resulting from the photometric analysis (see Sect.\,\ref{sec:activityAnalysis}). 

\subsection{GJ\,508.2}
\label{sec:GJ508_rv}
23 and 17 CARMENES VIS and NIR RV measurements, respectively, are available for the GJ\,508.2 M dwarf. The time series cover $\sim$3100\,days; they show a long-period variation with an amplitude that is consistent in both CARMENES channels. As before, we started by testing Keplerian fits using \texttt{Exo-Striker}. Although our monitoring does not yet sample a full cycle, the observed modulation points towards a system with an eccentric orbit. To estimate a lower limit to the period, we fitted the data with Keplerian orbits with increasing periods, computing the likelihood ($\mathcal{L}$) of the fit using \texttt{Exo-Striker}. All the other parameters, radial velocity semi-amplitude $K$, eccentricity $e$, argument of periastron $\omega$, and the time of periastron were left as free parameters. 

\begin{figure}[t]
\centering
\includegraphics[width=1.0\columnwidth]{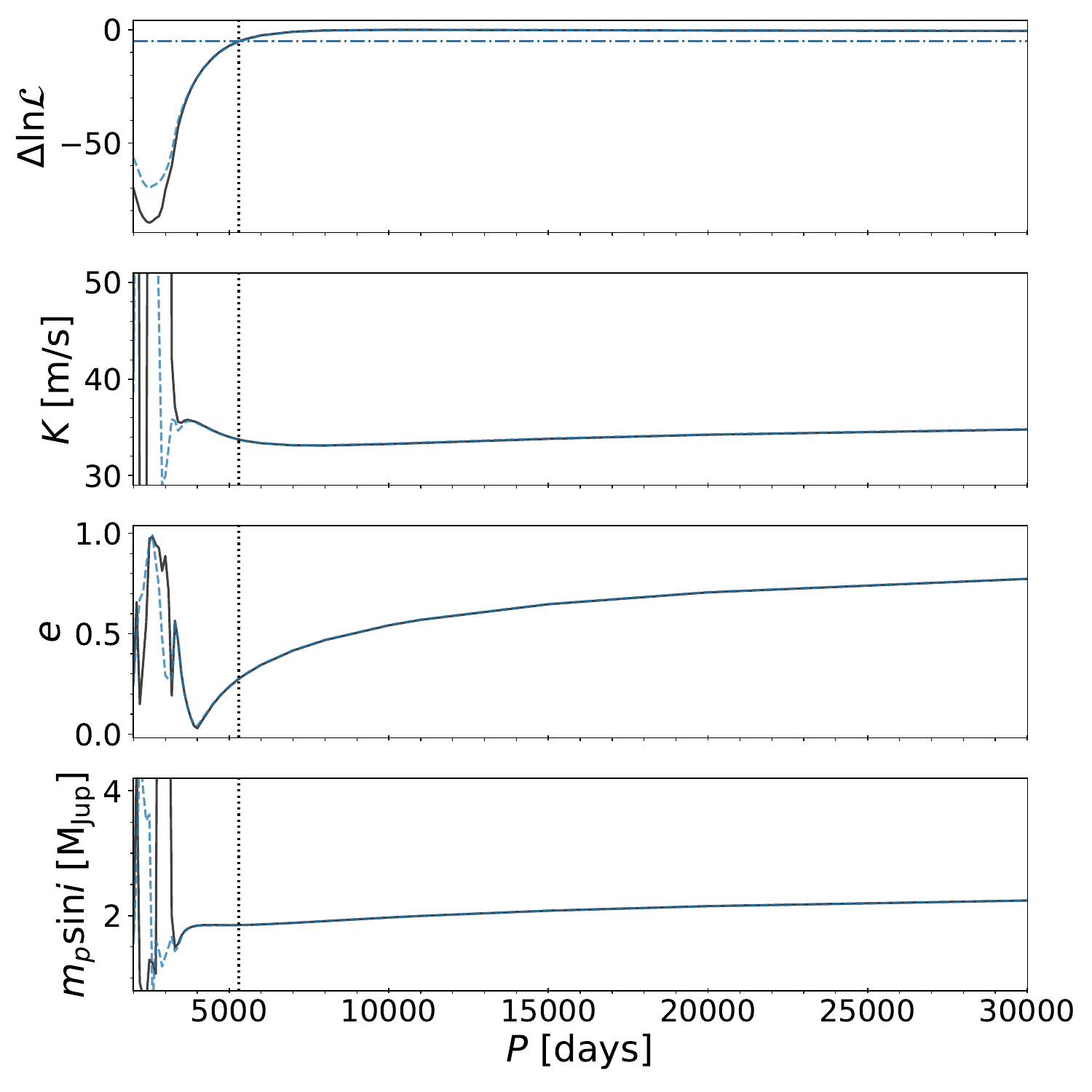}
\caption{Parameters of the fits to the RV time series of GJ\,508.2 as a function of the orbital period. The differential \lnL\, of the fit with respect to the maximum value, the fitted RV semi-amplitude and eccentricity, and the corresponding planet minimum mass are shown from top to bottom. Dahsed blue and solid black lines correspond, respectively, to fits considering only RVs from CARMENES VIS, and considering both VIS and NIR RVs simultaneously. The dot-dashed blue line in the top panel indicates the threshold corresponding to $\Delta$ \lnL=5, and the vertical dotted lines, the corresponding orbital period.}
\label{fig:pfix_gj508}
\end{figure}

\begin{figure}[t]
\centering
\includegraphics[width=1.0\columnwidth]{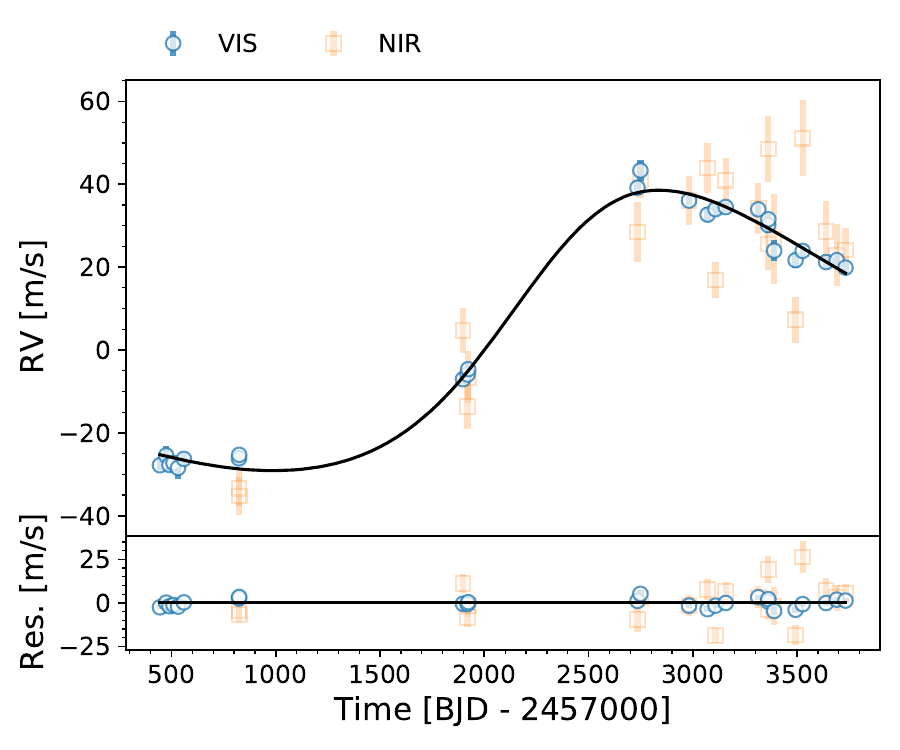}
\caption{RV curve of GJ\,508.2 with the best Keplerian fit shown as solid black line. RV measurements from the CARMENES VIS and NIR channels are depicted as blue circles and orange squares, respectively. Residuals are shown in the bottom panel.}
\label{fig:rv_gj508}
\end{figure}

Figure\,\ref{fig:pfix_gj508} shows the \lnL, the semi-amplitude, the eccentricity, and the corresponding minimum planet mass for each period tested (from top to bottom panels). From this figure, it is evident that any period above $\sim6000$\,days does not improve the fit in terms of \lnL, while shorter periods are clearly disfavoured. To estimate the lower limit of the orbital period, we put a threshold of 5 with respect to the best \lnL\, found, which results in a value of 5300\,days. This is marked as the dotted vertical line in Fig.\,\ref{fig:pfix_gj508}. We followed the same procedure using only the more precise VIS channel data, finding the same results as shown in the figure. Interestingly, the RV semi-amplitudes for the solutions above the threshold consistently take values $\sim$33\,\ms, while the orbital eccentricity correlates with the period and only a lower limit $e\gtrsim0.27$ can be set. This allows us to constrain the minimum mass of the planet candidate orbiting GJ\,508.2 to $\gtrsim1.85$\,\mjup, but below 2\,\mjup\, for orbital periods up to $\sim$40\,years. These lower limits for the parameters of this system are reported in the last column in Table\,\ref{tab:RVfit}. Fig.\,\ref{fig:rv_gj508} depicts the fit corresponding to the lower limit of the period for illustrative purposes.

\section{Constraints from astrometry}

\label{sec:astrometry}

The astrometric excess noise reported in the \textit{Gaia} third data release \citep[DR3,][]{Gaia2020,Gaia2023} as given in Table~\ref{tab:stars} is very significant for all the planet host stars analysed in this work. Under the assumption that this astrometric access noise is fully due to the one or two planets analysed in this work around each host star, one can constrain the missing orbital parameters  inclination and ascending node, or, if that is not feasible, at least to put a lower limit on the inclination and thus an upper limit on the planet mass.

Following the approach described in \cite{Morales2019}, we varied the orbital inclination and ascending node of the orbits and evaluated the configurations that resulted in an astrometric fit with an rms equal to the astrometric excess noise to within its errors. We estimated the \textit{Gaia} single measurement precision for each star in right ascension and declination from the formal errors on the final catalogue positions, taking the number of measurements for each into account. We also used the predicted individual timings of the \textit{Gaia} visits to the star and the corresponding scan angles according to the \textit{Gaia} attitude model provided by the \textit{Gaia} Observation Forecast Tool \footnote{\url{https://gaia.esac.esa.int/gost/}} relevant for \textit{Gaia} DR3. Finally, we used the given astrometric excess noise and its corresponding error to simulate abscissa residuals consistent with all constraints. We fitted an astrometric standard model to the simulated abscissa residuals, obtaining new residuals which can now be used for orbit fitting, taking the RV orbital parameters as well as their errors into account. For multiple systems, the orbits of both planets were computed simultaneously assuming that they have the same inclination and ascending node (there are not enough constraints to fit them separately).

For those cases where proper motion anomalies between \textit{Hipparcos} and \textit{Gaia} DR3 \citep{Kervella2022} were available (GJ\,463, GJ\,508.2 and GJ\,9773), we used these as additional constraints on the astrometric orbit following \cite{Sozzetti2023}. The computed three $\chi^{2}$ values (astrometric excess noise and proper motion anomaly in right ascension and declination) were weighted equally for the final likelihood. We used a nested sampling approach \citep{dynesty} for the fitting. While the formal errors on the primary constraints (astrometric excess noise and proper motion anomaly) are taken into account already in the computation of the $\chi^{2}$ value, this is not the case for the spectroscopic orbital parameters and the parallax, which also carry uncertainties which should be taken into account. We thus repeated the nested sampling 1000 times for each system while varying the spectroscopic orbital parameters and the parallax within their formal uncertainties. The final posteriors for inclination and node are shown in Appendix\,\ref{app:astrometry}, and the modes with their 1$\sigma$ confidence regions are given in Table\,\ref{tab:astrometry}.

With the assumption mentioned already above that the planets dominating the RV are also the ones  dominating the astrometry, this analysis provides masses of 0.78\aunc{0.04}{-0.27}\,\mjup\, and 0.77\aunc{0.04}{-0.26}\,\mjup\, for GJ\,3512 b and c, respectively, and masses of 3.02\aunc{0.14}{-0.41}\,\mjup\, and 2.56\aunc{0.14}{-0.35}\,\mjup\, for GJ\,317 b and c, respectively. The GJ\,9773 system hosts a planet with a mass of 1.43\aunc{0.52}{-0.08}\,\mjup\,, while assuming the lower limit period of GJ\,508.2 b we derive a mass of this planet of 6.0\aunc{0.7}{-1.2}\,\mjup, also well within the planetary mass domain. However, for this system, the period is not yet constrained, and hence the mass estimate is not as robust as in the other cases. Finally, for GJ\,463 b we derive a mass of 4.44\aunc{0.31}{-0.29}\,\mjup, while the value reported in \cite{Sozzetti2023} is 3.6$\pm$0.4\,\mjup, consistent to within about 2$\sigma$.

While this approach of using the \textit{Gaia} astrometric excess noise, the predicted timings and orientations of the \textit{Gaia} visits and the proper motion anomalies provides good constraints with small formal errors for the inclination and ascending node for most of the examined cases, there are some caveats to take into account when interpreting the results. First of all, as mentioned already, it is possible that the astrometric excess noise and the proper motion anomaly are dominated by hitherto unknown planets with considerable astrometric signals; these typically would have to be planets in larger orbits (so that they are noticeable or even dominate the astrometry, but not the radial velocities). In this case the provided values for the (absolute) inclinations should be regarded lower limits, while the derived masses should be considered upper limits to the real values. Second, one should also remember that we have essentially simulated our measurements, by generating \textit{Gaia} abscissa residuals which have the correct astrometric access noise, but which of course are only one possible representation of the real values and will not possess the correct properties in all details. For example, we have generated residuals which always have about the same size, with variations reflecting only measurement errors. However, the real measurements could have larger variations. This might potentially bias the astrometric orbits (which is the real orbit projected into the plane of the sky) to be more circular than in reality, which could potentially bias the inclinations to smaller values (but not too small, because then the astrometric excess noise would be larger than the measured one). These caveats will all be overcome with the release of epoch astrometry as part of the \textit{Gaia} DR4, when it should be possible to fully constrain astrometric orbits of Jupiter-mass planets in long period orbits around nearby stars, either from astrometry or alone or in combination with precise radial velocity datasets.

\section{Discussion}
\label{sec:discussion}
\subsection{Giant planets orbiting M dwarfs}
\label{subsec:planets}
Our previous analyses using CARMENES data concluded that giant planets orbiting M dwarfs are more abundant at larger orbital separations. \cite{Sabotta2021} and \cite{Ribas2023} found an occurrence rate of planets with a mass above 100\,M$_{\oplus}$ and an orbital period above 100 days of about 3\% for two CARMENES target subsamples, including 71 and 238 stars of the CARMENES survey, respectively. The planetary systems analysed here support this conclusion, as all of them formed beyond the present-day ice line \citep[$\sim$0.6\,au and $\sim$0.05\,au for M0 and M8-type stars, respectively, as estimated from][]{Ida2005}. A detailed analysis of the occurrence rates derived from CARMENES will be provided in subsequent complete data releases. However, we provide here newly updated estimates including the planets that we are reporting. 

To date, about 260 targets have more than 20 observations with CARMENES. For them, exoplanets with masses above 0.2\,\mjup\, as the ones reported here would have been detected because of their induced large RV variability, up to periods of about 10 years. Within this sample, the CARMENES survey has uncovered a total of six of such high-mass, long-period planets, namely, GJ\,3512 b and c, GJ\,9773 b, and GJ\,508.2 b in this work, and TZ\,Ari b and TYC\,2187-512-1 b in \cite{Quirrenbach2022}. A total of an additional 11 planets have also been reported in the 260-target sample and were analysed in other works: GJ\,317 b and c and GJ\,463 b (also analysed in this work), GJ\,876 b and c \citep{Trifonov2018}, GJ\,1148 b and c \citep{Trifonov2020b}, GJ\,179 b \citep{Howard2010}, GJ\,849 b and c \citep{Butler2006,Rosenthal2021}, and GJ\,649 b \citep{Johnson2010}. This yields a total of 17 planets with a mass above 0.2\,\mjup\ in the sample. A straightforward calculation indicates a global occurrence rate of this class of planets of at least $\sim$6.5\%, assuming they could be found with the current data for all stars in this CARMENES subsample. 

\begin{figure}
\centering
\includegraphics[width=0.5\textwidth]{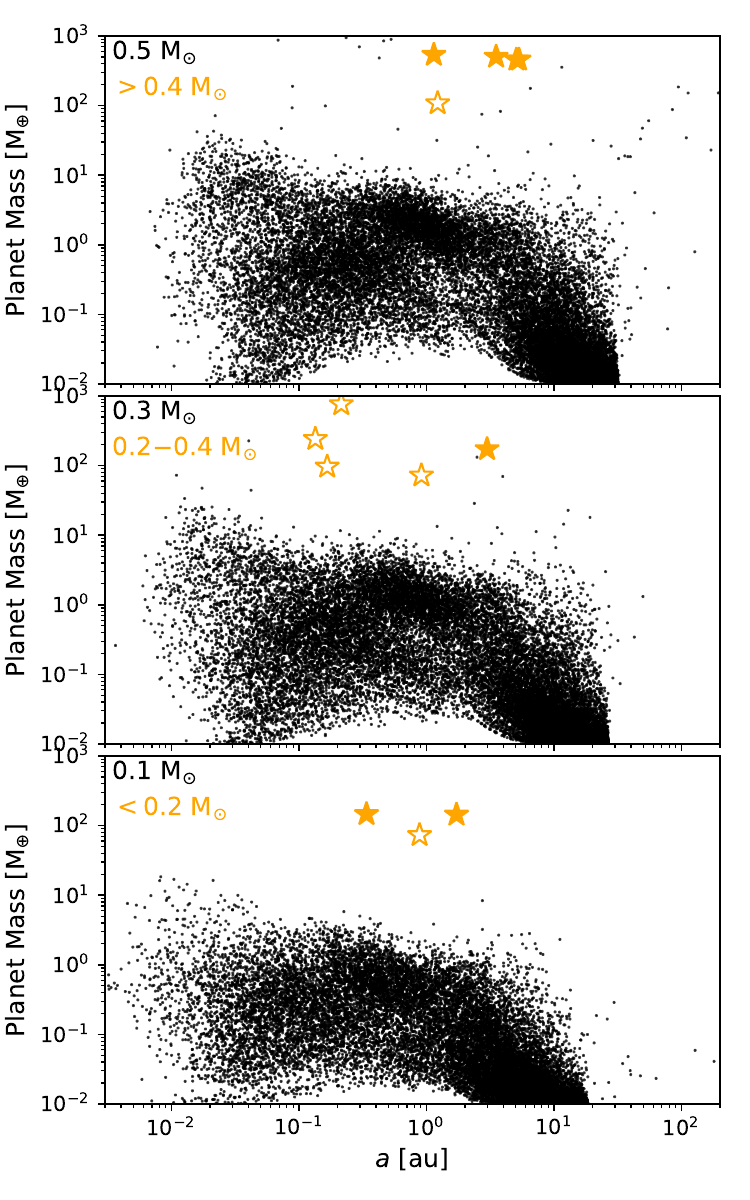}
\caption{Distribution of the mass of planets around M-dwarf stars as a function of their orbital semi-major axes obtained from the population synthesis models described in \cite{Burn2021}. The cases of host stars with masses of 0.5, 0.3, and 0.1\,M$_{\odot}$ are shown from the top to bottom panels, as representative of early-, mid-, and late-type M dwarfs. Planets in the CARMENES sample are illustrated as filled (this work) and empty \citep{Trifonov2018,Trifonov2020,Quirrenbach2022} symbols. They are represented on each panel in three host star mass bins as labelled.}
\label{fig:population}%
\end{figure}

This rather high occurrence rate is apparently inconsistent with planet population synthesis models around late-type stars based on planetesimal core/pebble accretion \citep{Mordasini2012,Emsenhuber2021}, which predict a smaller occurrence rate. Figure\,\ref{fig:population} shows the expected distribution of planets as a function of their orbital semi-major axes obtained with the Bern models as published in \cite{Burn2021}. The distribution is shown for three host star masses, decreasing from top to bottom, as proxies of early-, mid- and late-type stars. From these plots, we estimate that about $\sim$3\% of the simulated systems formed a gas-giant planet in the case of a 0.5\,M$_{\odot}$ host star at any separation. However, this number decreases to 0.4\% for 0.3\,M$_{\odot}$, while no gas-giant planets at all are formed around a 0.1\,M$_{\odot}$ host star. Forming Jupiter-mass planets for such small stars is more difficult because of their expected reduced proto-planetary disc mass \citep{Pascucci2016} when compared to Sun-like stars. An alternative scenario, proposed by \cite{Burn2021}, is that slower migration rates could result in the formation of more massive planets.

\begin{figure*}[t]
\centering
\includegraphics[width=0.49\textwidth]{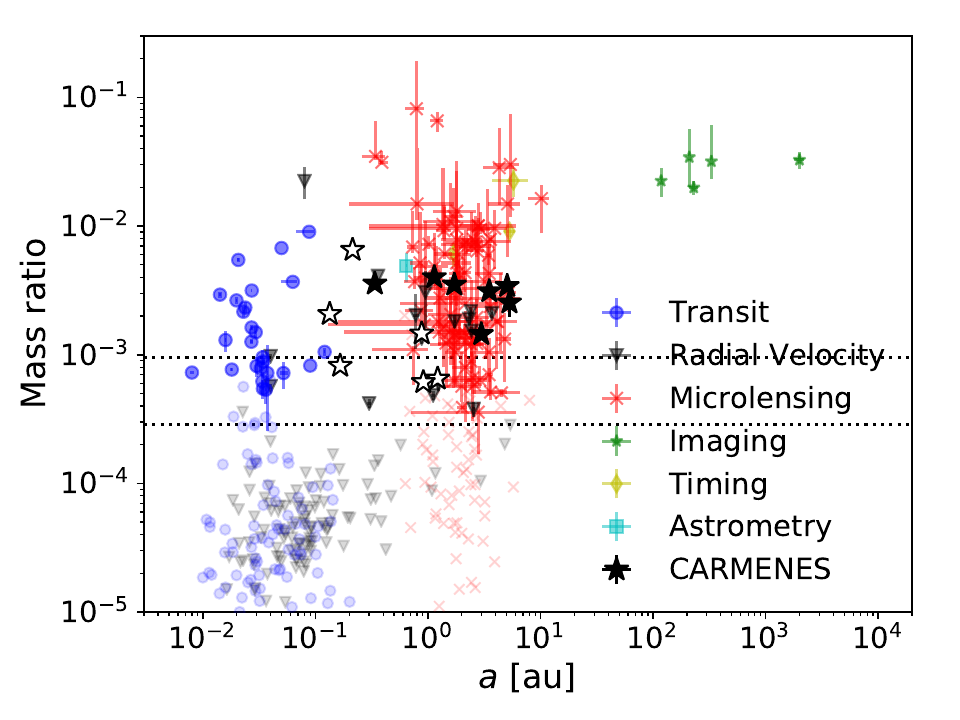}
\includegraphics[width=0.49\textwidth]{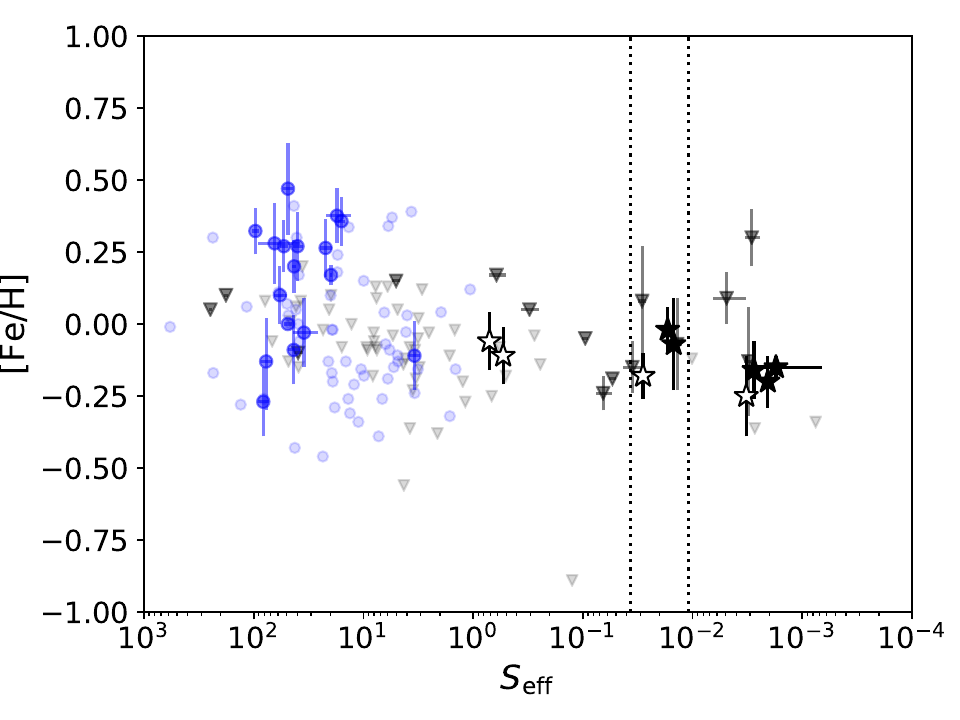}
\caption{Left: Planet-to-star mass ratio as a function of the planet semi-major axis for all planets orbiting M-dwarf stars. Darker coloured symbols correspond to planetary masses above 0.2\,\mjup. Horizontal dotted lines indicate the Jupiter-to-Sun and Saturn-to-Sun mass ratios for reference. Right: Host star metallicity as a function of the insolation at the exoplanet semi-major axis. For multi-planetary systems, only the innermost planet is depicted. Vertical dotted lines indicate current insolation values of Jupiter and Saturn for reference. In both panels, different detection techniques are shown as labelled. Star symbols depict planets in the CARMENES sample \citep[filled symbols, this work, and empty symbols,][]{Trifonov2018,Trifonov2020b,Quirrenbach2022}.}
\label{fig:massratio}
\end{figure*}

We compared the giant planets observed with CARMENES, either discovered or analysed \citep[this work and][]{Quirrenbach2022,Trifonov2018,Trifonov2020b}, with all the known systems around M dwarf stars listed in the NASA Exoplanets Archive\footnote{\url{https://exoplanetarchive.ipach.caltech.edu}}. According to this list, there are 160 planets with measured masses that orbit 149 M dwarf stars. A large fraction of them have been discovered by the microlensing technique ($\sim$60\%), while only 26 and 22 systems were found in RV and transiting planet surveys, respectively. The left panel in Fig.\,\ref{fig:massratio} illustrates the planet-to-stellar mass ratio as a function of the orbital semi-major axis. The new planet candidates announced here, GJ\,3512\,c, GJ\,9773\,b, and GJ\,508.2\,b, add to the low number of known gas-giant planets orbiting M-dwarf stars with precise dynamical mass determination from RV, which bridge the gap between the discoveries by the transit and the microlensing techniques. They show a range of values of the planet-to-star mass ratio from similar to the Sun-Saturn mass ratio to several times that of Sun-Jupiter, with a mean value of $\sim$0.003. A total of 6 out of the 21 RV planetary systems are multiple. This is comparable to the $\sim$29\% multiplicity fraction estimated from the same database for RV planets around more massive host stars ($>$0.6\,M$_{\odot}$).

\subsection{Metallicity of cold Jupiters}
\label{subsec:metallicity}
Many studies have reported that FGK host stars with giant planets tend to have higher metallicities, favouring the hypothesis of planet formation by core/pebble accretion in a proto-planetary disc with heavier elements \citep[see e.g.][for a review]{Fischer2005,Udry2007,Zhu2021}. In the particular case of M-dwarf host stars, \cite{Maldonado2020} concluded that the same trend is found. Interestingly, all M-dwarf host stars analysed in this work have subsolar metal content, according to the metallicities derived from the CARMENES spectra \citep{Marfil2021}. The right panel of Fig.\,\ref{fig:massratio} shows the metallicity of the M dwarf hosts stars as a function of the insolation of the planet for RV and transiting planetary systems. Only planets with reported stellar and planetary masses and metallicity determinations in the NASA Exoplanets Archive are depicted. The sample of such systems is composed of a total of 43 host stars when including the planets with CARMENES data listed above. An inspection of the plot shows a rather conspicuous difference between the metallicity of the giant planets that are in close-in with respect to those that reside in wide orbits around their host stars. Such a trend was also indicated by \cite{Kagetani2023}.

To quantitatively evaluate their difference, independently of observational biases, we counted the number of stellar hosts with metallicity above and below the solar value, and computed their ratio as $N($[Fe/H]$\ge0)/N($[Fe/H]$<0)$. The uncertainty of this ratio was estimated by simulating 10$^5$ random samples of the metallicity of the sample of host stars and taking into account the reported error bars or assuming an uncertainty of 0.2\,dex when error bars were not available. The results of this analysis are summarised in Table\,\ref{tab:metallicity}. We found a ratio of 0.64\aunc{0.24}{-0.55} for RV planets, while the value increases to 1.5$\pm$0.83 for transiting systems. As a further check, we performed a Kolmogorov-Smirnov test \citep{Kolmogorov1933,Smirnov1939} to compare the metallicities of the host stars with Jupiter-like planets detected by means of the transit and the radial velocity techniques. We obtained a $p$-value of 0.08, indicating that both samples are most likely coming from different populations. For comparison, we also computed the metal-rich to metal-poor ratio for more massive host stars ($>$0.6\,M$_{\odot}$) hosting giant planets with a mass $>$0.2\,\mjup. This resulted in a ratio of 1.4\aunc{0.27}{-0.15} and 2.99\aunc{0.34}{-0.57}, for RV and transiting planets, respectively. While these numbers should be taken with caution, given the heterogeneity of the metallicity values in the NASA exoplanets archive and the possible offsets between different techniques to measure the metallicity, their large difference may be pointing towards different formation or migration mechanisms and histories for the giant planets at wide distances of their M-dwarf host stars when compared with their close-in counterparts. Interestingly, \cite{Gan2025} reach the same conclusion by homogeneously measuring the metallicity of a limited sample of M-dwarf stars hosting 22 giant planets.

\begin{table}
\caption[]{Number of exoplanet host stars with metallicity determination.}
\label{tab:metallicity}
\begin{tabular}{rcccc}
\hline
\hline
\noalign{\smallskip}
[Fe/H]              & \multicolumn{2}{c}{Transits}         & \multicolumn{2}{c}{RV}  \\
range               & $<$0.2\,\mjup\,  & $\ge$0.2\,\mjup\, &  $<$0.2\,\mjup\, & $\ge$0.2\,\mjup\,\\
\noalign{\smallskip}
\hline
\noalign{\smallskip}
                    & \multicolumn{4}{c}{Host star mass $\le$0.6\,M$_{\odot}$}\\
\noalign{\smallskip}
\hline
\noalign{\smallskip}
 $\ge0$             & 25                     & 12                     & 17                     & 9   \\
 $<0$               & 43                     & 8                      & 40                     & 14  \\
 Ratio              & 0.58\aunc{0.12}{-0.07} & 1.5$\pm$0.83           & 0.42\aunc{0.16}{-0.26} & 0.64\aunc{0.24}{-0.55} \\
\noalign{\smallskip}
\hline
\noalign{\smallskip}
                    & \multicolumn{4}{c}{Host star mass $>$0.6\,M$_{\odot}$}\\
\noalign{\smallskip}
\hline
\noalign{\smallskip}
 $\ge0$             & 165                    & 416                    & 44               & 210   \\
 $<0$               & 95                     & 139                    & 34               & 150  \\
 Ratio              & 1.74\aunc{0.43}{-0.32} & 2.99\aunc{0.34}{-0.57} & 1.29$\pm$0.14    & 1.40\aunc{0.27}{-0.15} \\
\noalign{\smallskip}
\hline
\end{tabular}

\textbf{Notes:} Metallicity values reported in this work and retrieved from the NASA Exoplanets Archive are used to obtain these numbers.
\end{table}

\section{Conclusions}
\label{sec:conclusions}
We report the properties of the outer giant planet orbiting the very-low-mass star GJ\,3512, which hosts two almost identical planets with masses of $\sim$0.46\,\mjup\, orbiting at distances of 0.46\,au and 1.73\,au. CARMENES and IRD high-resolution spectroscopic observations in the visual and NIR channels have been used for the analysis. GJ\,3512\,c is the third giant planet confirmed to orbit a late-type star with a mass $<$0.15\,M$_{\odot}$. Furthermore, we announce the discovery of two new Jupiter-like planets orbiting M-dwarf stars within the CARMENES survey. These are GJ\,9773\,b, a half-Jupiter minimum mass planet, and the candidate GJ\,508.2\,b, a $\sim$2\,\mjup\, planet; they orbit mid- and early-M-type stars, respectively, with periods of several years. For GJ\,9773\,b, we estimate an absolute mass of 1.43\aunc{0.52}{-0.08}\,\mjup\ based on the proper motion anomaly. We have also determined the rotation period of the star GJ\,9773 and offer a new determination for GJ\,508.2. Furthermore, we have refined the properties of the planetary systems orbiting GJ\,317 and GJ\,463, and inferred an absolute mass of 4.44\aunc{0.31}{0.29}\,\mjup\, for GJ\,463\,b .

The systems we analysed confirm that the occurrence rate of Jupiter-like planets is higher than expected from theoretical exoplanet formation models in the core or pebble accretion paradigm, which points to alternative scenarios such as gravitational instability. The lower ratio of high-metallicity to low-metallicity stars that host long-period Jupiter-like planets compared to that of the stars with close-in giant planets may well point towards different formation or migration mechanisms. The current RV surveys targeting M-dwarf stars, such as those conducted with CARMENES, SPIRou \citep{Moutou2023}, and IRD \citep{Harakawa2022}, are helping to increase the known population of such systems. The next data release from the \textit{Gaia} mission will also resolve the astrometric orbit of many more giant planets around M-dwarf stars, and provide a larger and more homogeneous sample of systems. This will be key in elucidating their formation mechanisms and their dynamical evolution. Interestingly, several of these planets will also be accessible from future space facilities such as the \textit{Roman} space telescope, the \textit{Habitable Worlds Observatory}, and the \textit{LIFE} interferometry mission. The study of their thermal emission and their reflection properties will provide more insight into their nature.

\section*{Data availability}
Tables \ref{tab:RVs_GJ3512}, \ref{tab:RVs_GJ317}, \ref{tab:RVs_GJ463}, \ref{tab:RVs_GJ9773}, \ref{tab:RVs_GJ508}, \ref{tab:LC_GJ9773}, and \ref{tab:LC_GJ508} are only available in electronic form at the CDS via anonymous ftp to \url{cdsarc.u-strasbg.fr (130.79.128.5)} or via \url{http://cdsweb.u-strasbg.fr/cgi-bin/qcat?}.

\begin{acknowledgements}
This publication was based on observations collected under the CARMENES Legacy+ project. CARMENES is an instrument at the Centro Astron\'omico Hispano en Andaluc\'ia (CAHA) at Calar Alto (Almer\'{\i}a, Spain), operated jointly by the Junta de Andaluc\'ia and the Instituto de Astrof\'isica de Andaluc\'ia (CSIC). The authors wish to express their sincere thanks to all members of the Calar Alto staff for their expert support of the instrument and telescope operation. CARMENES was funded by the Max-Planck-Gesellschaft (MPG), the Consejo Superior de Investigaciones Cient\'{\i}ficas (CSIC), the Ministerio de Econom\'ia y Competitividad (MINECO) and the European Regional Development Fund (ERDF) through projects FICTS-2011-02, ICTS-2017-07-CAHA-4, and CAHA16-CE-3978, and the members of the CARMENES Consortium (Max-Planck-Institut f\"ur Astronomie, Instituto de Astrof\'{\i}sica de Andaluc\'{\i}a, Landessternwarte K\"onigstuhl, Institut de Ci\`encies de l'Espai, Institut f\"ur Astrophysik G\"ottingen, Universidad Complutense de Madrid, Th\"uringer Landessternwarte Tautenburg, Instituto de Astrof\'{\i}sica de Canarias, Hamburger Sternwarte, Centro de Astrobiolog\'{\i}a and Centro Astron\'omico Hispano-Alem\'an), with additional contributions by the MINECO, the Deutsche Forschungsgemeinschaft (DFG) through the Major Research Instrumentation Programme and Research Unit FOR2544 ``Blue Planets around Red Stars'', the Klaus Tschira Stiftung, the states of Baden-W\"urttemberg and Niedersachsen, and by the Junta de Andaluc\'{\i}a. This work has made use of observations from the Las Cumbres Observatory global telescope network and of data from the European Space Agency (ESA) mission {\it Gaia} (\url{https://www.cosmos.esa.int/gaia}), processed by the {\it Gaia} Data Processing and Analysis Consortium (DPAC, \url{https://www.cosmos.esa.int/web/gaia/dpac/consortium}). Funding for the DPAC has been provided by national institutions, in particular the institutions participating in the {\it Gaia} Multilateral Agreement. We acknowledge financial support from the Agencia Estatal de Investigaci\'on (AEI/10.13039/501100011033) of the Ministerio de Ciencia e Innovaci\'on and the ERDF ``A way of making Europe'' through projects PID2020-120375GB-I00, PID2021-125627OB-C31, PID2022-137241NB-C4[1:4], CNS2022-136050, and the Centre of Excellence ``Mar\'ia de Maeztu'' and ``Severo Ochoa'' awards to the Institut de Ci\`encies de l'Espai (CEX2020-001058-M), Instituto de Astrof\'isica de Canarias (CEX2019-000920-S), and Instituto de Astrof\'isica de Andaluc\'ia (CEX2021-001131-S). This work was also funded by the European Research Council (ERC) under the European Union’s Horizon Europe  programme (ERC Advanced Grant SPOTLESS; no. 101140786), the DFG priority program SPP 1992 ``Exploring the Diversity of Extrasolar Planets'' through grants JE 701/5-1 and RE 1664/20-1, the Secretaria d'Universitats i Recerca del Departament d'Empresa i Coneixement de la Generalitat de Catalunya and the Ag\`encia de Gesti\'o d’Ajuts Universitaris i de Recerca of the Generalitat de Catalunya (SGR 01526/2021), with additional funding from the European FEDER/ERF funds, \emph{L'FSE inverteix en el teu futur}, from the Generalitat de Catalunya/CERCA programme, and the Israel Science Foundation through grant 1404/22. We thank T. Hirano and the other IRD-SSP members because of their contributions on the IRD observations and the data analysis, which greatly helped the progress of the IRD-SSP survey. The development and operation of IRD were supported by JSPS KAKENHI Grant Numbers 18H05442, 15H02063, and 22000005, and the Astrobiology Center (ABC) of NINS. This research is based in part on data collected at the Subaru Telescope, which is operated by the National Astronomical Observatory of Japan. We are honoured and grateful for the opportunity of observing the Universe from Maunakea, which has cultural, historical, and natural significance in Hawaii. Part of the data analysis was carried out on the Multi-wavelength Data Analysis System operated by the Astronomy Data Center (ADC), National Astronomical Observatory of Japan. Data reduction of IRD raw data was performed with IRAF, which is distributed by the National Optical Astronomy Observatories, which is operated by the Association of Universities for Research in Astronomy, Inc. (AURA) under cooperative agreement with the National Science Foundation. We acknowledge the anonymous referee for their useful comments to improve the manuscript.
\end{acknowledgements}

\bibliographystyle{aa} 
\bibliography{bibtex} 

\begin{appendix}
\section{Spectroscopy time series}
\label{app:RVdata}
The CARMENES and IRD RV measurements of GJ\,3512, GJ\,317, GJ\,463, GJ\,9773, and GJ\,508.2 used in this work are listed in Tables\,\ref{tab:RVs_GJ3512}, \ref{tab:RVs_GJ317}, \ref{tab:RVs_GJ463}, \ref{tab:RVs_GJ9773}, and \ref{tab:RVs_GJ508}, respectively. Table A.6, only available in electronic form, provides the time series of the spectroscopic indices used for the analysis in Section\,\ref{sec:activity_spectra}.

\begin{table}[H]
\begin{center}
\caption[]{GJ\,3512 CARMENES and IRD radial velocity measurements.}
\label{tab:RVs_GJ3512}
\begin{tabular}{rcc}
\hline
\hline
\noalign{\smallskip}
               &  RV             &              \\
BJD            &  (m\,s$^{-1}$)  & Instrument   \\
\noalign{\smallskip}
\hline                               
\noalign{\smallskip}
2457752.403057 & --96.8 $\pm$ 1.8 & VIS \\
2457755.382612 & --99.6 $\pm$ 2.7 & VIS \\
2457761.473134 & --101.5 $\pm$ 2.6 & VIS \\
2457788.499932 & --53.1 $\pm$ 3.4 & VIS \\
2457815.466089 & --21.1 $\pm$ 1.9 & VIS \\
2457823.566145 & --13.5 $\pm$ 2.0 & VIS \\
\hline
\end{tabular}
\end{center}
\end{table}
\addtocounter{table}{-1}

\begin{table}[H]
\begin{center}
\caption[]{Continued.}
\begin{tabular}{rcc}
\hline
\hline
\noalign{\smallskip}
               &  RV             &              \\
BJD            &  m\,s$^{-1}$    & Instrument   \\
\noalign{\smallskip}
\hline                               
\noalign{\smallskip}
2458143.411122 & --22.8 $\pm$ 1.9 & VIS \\
2458149.536033 & --48.8 $\pm$ 3.0 & VIS \\
2458164.388712 & --97.1 $\pm$ 2.2 & VIS \\
2458166.474616 & --98.1 $\pm$ 1.9 & VIS \\
2458167.474075 & --100.0 $\pm$ 2.5 & VIS \\
2458169.592361 & --96.5 $\pm$ 2.1 & VIS \\
\hline
\end{tabular}
\end{center}
\end{table}
\addtocounter{table}{-1}

\begin{table}[H]
\begin{center}
\caption[]{Continued.}
\begin{tabular}{rcc}
\hline
\hline
\noalign{\smallskip}
               &  RV             &              \\
BJD            &  m\,s$^{-1}$    & Instrument   \\
\noalign{\smallskip}
\hline                               
\noalign{\smallskip}
2458458.633046 & 29.5 $\pm$ 1.5 & VIS \\
2458474.631985 & 38.6 $\pm$ 1.9 & VIS \\
2458485.608055 & 42.8 $\pm$ 1.9 & VIS \\
2458492.572300 & 47.7 $\pm$ 2.0 & VIS \\
2458499.571373 & 55.4 $\pm$ 2.4 & VIS \\
2458510.646602 & 65.8 $\pm$ 2.2 & VIS \\
\hline
\end{tabular}
\end{center}
\end{table}
\addtocounter{table}{-1}

\begin{table}[H]
\begin{center}
\caption[]{Continued.}
\begin{tabular}{rcc}
\hline
\hline
\noalign{\smallskip}
               &  RV             &              \\
BJD            &  m\,s$^{-1}$    & Instrument   \\
\noalign{\smallskip}
\hline                               
\noalign{\smallskip}
2459360.424345 & 39.8 $\pm$ 3.5 & VIS \\
2459372.376581 & --27.4 $\pm$ 1.8 & VIS \\
2459465.680212 & 58.9 $\pm$ 4.0 & VIS \\
2459475.688495 & 55.4 $\pm$ 2.5 & VIS \\
2459485.699582 & 63.3 $\pm$ 3.0 & VIS \\
2459501.688150 & 74.2 $\pm$ 2.8 & VIS \\
\hline
\end{tabular}
\end{center}
\end{table}
\addtocounter{table}{-1}

\begin{table}[H]
\begin{center}
\caption[]{Continued.}
\begin{tabular}{rcc}
\hline
\hline
\noalign{\smallskip}
               &  RV             &              \\
BJD            &  m\,s$^{-1}$    & Instrument   \\
\noalign{\smallskip}
\hline                               
\noalign{\smallskip}
2458073.721836 & 17.1 $\pm$ 8.6 & NIR \\
2458078.713052 & 34.0 $\pm$ 4.8 & NIR \\
2458084.692131 & 30.1 $\pm$ 5.4 & NIR \\
2458091.670869 & 30.4 $\pm$ 4.7 & NIR \\
2458093.644757 & 41.3 $\pm$ 4.5 & NIR \\
2458095.672439 & 37.0 $\pm$ 5.3 & NIR \\
\hline
\end{tabular}
\end{center}
\end{table}
\addtocounter{table}{-1}

\begin{table}[H]
\begin{center}
\caption[]{Continued.}
\begin{tabular}{rcc}
\hline
\hline
\noalign{\smallskip}
               &  RV             &              \\
BJD            &  m\,s$^{-1}$    & Instrument   \\
\noalign{\smallskip}
\hline                               
\noalign{\smallskip}
2458263.414442 & 16.6 $\pm$ 9.2 & NIR \\
2458264.404887 & 22.1 $\pm$ 7.4 & NIR \\
2458265.422043 & 23.2 $\pm$ 8.1 & NIR \\
2458266.406024 & 23.5 $\pm$ 7.1 & NIR \\
2458270.366495 & 23.5 $\pm$ 7.1 & NIR \\
2458271.388925 & 29.4 $\pm$ 7.5 & NIR \\
\hline
\end{tabular}
\end{center}
\end{table}
\addtocounter{table}{-1}

\begin{table}[H]
\begin{center}
\caption[]{Continued.}
\begin{tabular}{rcc}
\hline
\hline
\noalign{\smallskip}
               &  RV             &              \\
BJD            &  m\,s$^{-1}$    & Instrument   \\
\noalign{\smallskip}
\hline                               
\noalign{\smallskip}
2458801.692983 & --19.9 $\pm$ 9.8 & NIR \\
2458808.574528 & --41 $\pm$ 16 & NIR \\
2458815.635880 & --5 $\pm$ 14 & NIR \\
2458827.720635 & 34 $\pm$ 14 & NIR \\
2458835.681801 & 20 $\pm$ 11 & NIR \\
2458844.659009 & 30 $\pm$ 22 & NIR \\
\hline
\end{tabular}
\end{center}
\end{table}
\addtocounter{table}{-1}

\begin{table}[H]
\begin{center}
\caption[]{Continued.}
\begin{tabular}{rcc}
\hline
\hline
\noalign{\smallskip}
               &  RV             &              \\
BJD            &  m\,s$^{-1}$    & Instrument   \\
\noalign{\smallskip}
\hline                               
\noalign{\smallskip}
2460424.519725 & --80 $\pm$ 13 & NIR \\
2460579.688919 & --6.8 $\pm$ 6.7 & NIR \\
2460621.576685 & --61.8 $\pm$ 7.1 & NIR \\
\hline
2458779.141232 & --89.4 $\pm$ 3.5 & IRD \\
2458836.101612 & 6.5 $\pm$ 3.4 & IRD \\
2458880.810264 & 39.7 $\pm$ 4.1 & IRD \\
\hline
\end{tabular}
\end{center}
\end{table}

\begin{table}[H]
\begin{center}
\caption[]{GJ\,317 CARMENES radial velocity measurements.}
\label{tab:RVs_GJ317}
\begin{tabular}{rcc}
\hline
\hline
\noalign{\smallskip}
               &  RV            &              \\
BJD            &  m\,s$^{-1}$   & Instrument   \\
\noalign{\smallskip}
\hline                                    
\noalign{\smallskip}
2458093.666955 & --12.18 $\pm$ 0.94 & VIS \\
2458105.638951 & --19.7 $\pm$ 1.3 & VIS \\
2458110.624822 & --27.7 $\pm$ 1.3 & VIS \\
2458117.622890 & --30.0 $\pm$ 1.2 & VIS \\
2458122.597913 & --33.7 $\pm$ 1.4 & VIS \\
2458134.616968 & --40.0 $\pm$ 1.4 & VIS \\
2458139.555553 & --43.5 $\pm$ 1.4 & VIS \\
2458166.520357 & --52.5 $\pm$ 1.3 & VIS \\
2458188.416466 & --62.9 $\pm$ 2.2 & VIS \\
2458215.338093 & --69.7 $\pm$ 1.3 & VIS \\
2458449.690747 & --18.6 $\pm$ 2.7 & VIS \\
2458469.639265 & --4.1 $\pm$ 1.6 & VIS \\
2458490.588040 & 5.5 $\pm$ 1.1 & VIS \\
\noalign{\smallskip}
\hline
\end{tabular}
\end{center}
\end{table}
\addtocounter{table}{-1}

\begin{table}[H]
\begin{center}
\caption[]{Continued.}
\begin{tabular}{rcc}
\hline
\hline
\noalign{\smallskip}
               &  RV             &              \\
BJD            &  m\,s$^{-1}$    & Instrument   \\
\noalign{\smallskip}
\hline                               
\noalign{\smallskip}
2458511.519175 & 20.4 $\pm$ 2.4 & VIS \\
2458532.495504 & 41.2 $\pm$ 1.3 & VIS \\
2458554.411519 & 45.8 $\pm$ 1.3 & VIS \\
2458587.359534 & 63.5 $\pm$ 1.8 & VIS \\
2458786.728120 & --2.0 $\pm$ 1.8 & VIS \\
2458816.632507 & --14.3 $\pm$ 1.3 & VIS \\
\hline
\end{tabular}
\end{center}
\end{table}
\addtocounter{table}{-1}

\begin{table}[H]
\begin{center}
\caption[]{Continued.}
\begin{tabular}{rcc}
\hline
\hline
\noalign{\smallskip}
               &  RV             &              \\
BJD            &  m\,s$^{-1}$    & Instrument   \\
\noalign{\smallskip}
\hline                               
\noalign{\smallskip}
2460658.624908 & 90.1 $\pm$ 5.9 & NIR \\
2460688.585755 & 116 $\pm$ 11 & NIR \\
\noalign{\smallskip}
\hline
\end{tabular}
\end{center}
\end{table}

\begin{table}[H]
\begin{center}
\caption[]{GJ\,463 CARMENES radial velocity measurements.}
\label{tab:RVs_GJ463}
\begin{tabular}{rcc}
\hline
\hline
\noalign{\smallskip}
               &  RV          &              \\
BJD            &  m\,s$^{-1}$ & Instrument   \\
\noalign{\smallskip}
\hline              
\noalign{\smallskip}
2457449.690491 & 31.3 $\pm$ 3.4 & VIS \\
2457476.560499 & 28.0 $\pm$ 4.1 & VIS \\
2457492.499005 & 34.6 $\pm$ 4.0 & VIS \\
2457510.481961 & 34.8 $\pm$ 3.9 & VIS \\
2457557.432086 & 20.7 $\pm$ 4.2 & VIS \\
2457822.564107 & 8.1 $\pm$ 3.3 & VIS \\
\hline
\end{tabular}
\end{center}
\end{table}
\addtocounter{table}{-1}

\begin{table}[H]
\begin{center}
\caption[]{Continued.}
\begin{tabular}{rcc}
\hline
\hline
\noalign{\smallskip}
               &  RV             &              \\
BJD            &  m\,s$^{-1}$    & Instrument   \\
\noalign{\smallskip}
\hline                               
\noalign{\smallskip}
2457968.367271 & 2.8 $\pm$ 3.9 & VIS \\
2457969.376468 & 0.5 $\pm$ 3.7 & VIS \\
2457970.384874 & 0.0 $\pm$ 3.7 & VIS \\
2457971.364378 & 2.5 $\pm$ 3.8 & VIS \\
2457972.360671 & 0.4 $\pm$ 4.4 & VIS \\
2457974.352960 & 5.0 $\pm$ 3.7 & VIS \\
\hline
\end{tabular}
\end{center}
\end{table}
\addtocounter{table}{-1}

\begin{table}[H]
\begin{center}
\caption[]{Continued.}
\begin{tabular}{rcc}
\hline
\hline
\noalign{\smallskip}
               &  RV             &              \\
BJD            &  m\,s$^{-1}$    & Instrument   \\
\noalign{\smallskip}
\hline                               
\noalign{\smallskip}
2458544.628609 & --22.4 $\pm$ 3.8 & VIS \\
2458554.563080 & --15.1 $\pm$ 3.0 & VIS \\
2458564.602681 & --16.0 $\pm$ 3.6 & VIS \\
2458587.583490 & --20.4 $\pm$ 3.0 & VIS \\
2458597.535912 & --17.2 $\pm$ 3.9 & VIS \\
2458608.542063 & --20.9 $\pm$ 3.4 & VIS \\
\hline
\end{tabular}
\end{center}
\end{table}
\addtocounter{table}{-1}

\begin{table}[H]
\begin{center}
\caption[]{Continued.}
\begin{tabular}{rcc}
\hline
\hline
\noalign{\smallskip}
               &  RV             &              \\
BJD            &  m\,s$^{-1}$    & Instrument   \\
\noalign{\smallskip}
\hline                               
\noalign{\smallskip}
2457909.461893 & 14.3 $\pm$ 6.0 & NIR \\
2457910.409572 & 9.8 $\pm$ 4.1 & NIR \\
2457911.364648 & 12.8 $\pm$ 4.9 & NIR \\
2457912.410432 & 8.2 $\pm$ 4.8 & NIR \\
2457914.453125 & 16.7 $\pm$ 4.3 & NIR \\
2457915.424742 & 9.3 $\pm$ 5.5 & NIR \\
\hline
\end{tabular}
\end{center}
\end{table}
\addtocounter{table}{-1}

\begin{table}[H]
\begin{center}
\caption[]{Continued.}
\begin{tabular}{rcc}
\hline
\hline
\noalign{\smallskip}
               &  RV             &              \\
BJD            &  m\,s$^{-1}$    & Instrument   \\
\noalign{\smallskip}
\hline                               
\noalign{\smallskip}
2458139.700248 & --2.4 $\pm$ 5.9 & NIR \\
2458140.696842 & --6.1 $\pm$ 4.6 & NIR \\
2458141.701508 & 3.4 $\pm$ 4.5 & NIR \\
2458142.687977 & 6.9 $\pm$ 4.7 & NIR \\
2458161.631025 & --19 $\pm$ 15 & NIR \\
2458161.657693 & --7.2 $\pm$ 6.6 & NIR \\
\hline
\end{tabular}
\end{center}
\end{table}

\begin{table}[H]
\begin{center}
\caption[]{GJ\,9773 CARMENES radial velocity measurements.}
\label{tab:RVs_GJ9773}
\begin{tabular}{rcc}
\hline
\hline
\noalign{\smallskip}
               &  RV            &              \\
BJD            &  m\,s$^{-1}$   & Instrument   \\
\noalign{\smallskip}
\hline      
\noalign{\smallskip}
2457571.646992 & --10.0 $\pm$ 1.4 & VIS \\
2457573.665074 & --13.3 $\pm$ 2.1 & VIS \\
2457603.598854 & --7.2 $\pm$ 3.5 & VIS \\
2457628.540999 & --7.7 $\pm$ 1.9 & VIS \\
2457629.531752 & --10.0 $\pm$ 1.8 & VIS \\
2457690.384828 & 1.7 $\pm$ 3.0 & VIS \\
\hline
\end{tabular}
\end{center}
\end{table}
\addtocounter{table}{-1}

\begin{table}[H]
\begin{center}
\caption[]{Continued.}
\begin{tabular}{rcc}
\hline
\hline
\noalign{\smallskip}
               &  RV             &              \\
BJD            &  m\,s$^{-1}$    & Instrument   \\
\noalign{\smallskip}
\hline                               
\noalign{\smallskip}
2458454.361283 & 6.9 $\pm$ 1.4 & VIS \\
2458467.348932 & 9.1 $\pm$ 2.3 & VIS \\
2458475.322980 & 6.0 $\pm$ 1.9 & VIS \\
2458483.275644 & 7.2 $\pm$ 1.6 & VIS \\
2458496.260188 & 7.7 $\pm$ 1.3 & VIS \\
2458608.667665 & 7.3 $\pm$ 1.5 & VIS \\
\hline
\end{tabular}
\end{center}
\end{table}
\addtocounter{table}{-1}

\begin{table}[H]
\begin{center}
\caption[]{Continued.}
\begin{tabular}{rcc}
\hline
\hline
\noalign{\smallskip}
               &  RV             &              \\
BJD            &  m\,s$^{-1}$    & Instrument   \\
\noalign{\smallskip}
\hline                               
\noalign{\smallskip}
2460515.662486 & --10.6 $\pm$ 1.9 & VIS \\
2460536.594471 & --6.9 $\pm$ 2.0 & VIS \\
2460550.490488 & --9.4 $\pm$ 1.2 & VIS \\
2460562.486414 & --6.5 $\pm$ 1.5 & VIS \\
2460572.464864 & --10.8 $\pm$ 1.4 & VIS \\
2460582.381256 & --11.1 $\pm$ 1.6 & VIS \\
\hline
\end{tabular}
\end{center}
\end{table}
\addtocounter{table}{-1}

\begin{table}[H]
\begin{center}
\caption[]{Continued.}
\begin{tabular}{rcc}
\hline
\hline
\noalign{\smallskip}
               &  RV             &              \\
BJD            &  m\,s$^{-1}$    & Instrument   \\
\noalign{\smallskip}
\hline                               
\noalign{\smallskip}
2458411.498965 & 8.9 $\pm$ 9.9 & NIR \\
2458433.395506 & --10.9 $\pm$ 6.6 & NIR \\
2458447.404738 & --6 $\pm$ 14 & NIR \\
2458454.360733 & 8.1 $\pm$ 6.2 & NIR \\
2458467.349506 & 4 $\pm$ 11 & NIR \\
2458475.321667 & 16.0 $\pm$ 7.2 & NIR \\
\hline
\end{tabular}
\end{center}
\end{table}
\addtocounter{table}{-1}

\begin{table}[H]
\begin{center}
\caption[]{Continued.}
\begin{tabular}{rcc}
\hline
\hline
\noalign{\smallskip}
               &  RV             &              \\
BJD            &  m\,s$^{-1}$    & Instrument   \\
\noalign{\smallskip}
\hline                               
\noalign{\smallskip}
2460503.575230 & --6.8 $\pm$ 7.0 & NIR \\
2460515.662284 & --8.9 $\pm$ 5.4 & NIR \\
2460526.561430 & --16.2 $\pm$ 6.1 & NIR \\
2460536.594498 & --2.9 $\pm$ 7.0 & NIR \\
2460550.490017 & --3.2 $\pm$ 3.5 & NIR \\
2460562.486724 & --15.0 $\pm$ 8.7 & NIR \\
2460572.464804 & --12.0 $\pm$ 4.9 & NIR \\
2460582.382082 & --17.0 $\pm$ 6.1 & NIR \\
2460592.445773 & 11 $\pm$ 12 & NIR \\
2460621.360151 & --11.4 $\pm$ 4.3 & NIR \\
\noalign{\smallskip}
\hline
\end{tabular}
\end{center}
\end{table}

\begin{table}[h]
\begin{center}
\caption[]{GJ\,508.2 CARMENES radial velocity measurements.}
\label{tab:RVs_GJ508}
\begin{tabular}{rcc}
\hline
\hline
\noalign{\smallskip}
              &  RV           &              \\
BJD           &  m\,s$^{-1}$  & Instrument   \\
\noalign{\smallskip}
\hline  
\noalign{\smallskip}
2457444.665750 & --36.8 $\pm$ 1.2 & VIS \\
2457474.547304 & --34.4 $\pm$ 2.2 & VIS \\
2457490.526201 & --36.7 $\pm$ 1.8 & VIS \\
2457509.494132 & --36.2 $\pm$ 1.6 & VIS \\
2457531.402648 & --37.4 $\pm$ 2.7 & VIS \\
2457559.454673 & --35.2 $\pm$ 1.5 & VIS \\
2457823.528421 & --35.1 $\pm$ 1.6 & VIS \\
2457824.649987 & --34.2 $\pm$ 1.2 & VIS \\
2458897.665336 & --16.0 $\pm$ 1.1 & VIS \\
2458919.597608 & --14.9 $\pm$ 1.3 & VIS \\
2458922.622462 & --13.6 $\pm$ 1.5 & VIS \\
2459734.549112 & 30.2 $\pm$ 1.7 & VIS \\
2459748.431925 & 34.4 $\pm$ 2.5 & VIS \\
2459981.591301 & 27.2 $\pm$ 1.2 & VIS \\
2460070.581953 & 23.7 $\pm$ 1.9 & VIS \\
2460108.426309 & 25.1 $\pm$ 1.3 & VIS \\
2460157.363021 & 25.6 $\pm$ 1.1 & VIS \\
2460313.716112 & 25.0 $\pm$ 1.4 & VIS \\
2460360.668123 & 21.2 $\pm$ 2.0 & VIS \\
2460361.673099 & 22.6 $\pm$ 1.2 & VIS \\
2460389.700836 & 15.0 $\pm$ 2.5 & VIS \\
2460492.435151 & 12.8 $\pm$ 1.8 & VIS \\
2460526.367748 & 15.0 $\pm$ 1.5 & VIS \\
2460638.728074 & 12.3 $\pm$ 1.6 & VIS \\
2460689.755505 & 12.7 $\pm$ 1.3 & VIS \\
2460731.664176 & 10.9 $\pm$ 1.4 & VIS \\
\hline
2457823.528078 & --42.3 $\pm$ 4.2 & NIR \\
2457824.651332 & --44.2 $\pm$ 4.6 & NIR \\
2458897.664813 & --4.3 $\pm$ 5.4 & NIR \\
2458919.597566 & --22.5 $\pm$ 5.5 & NIR \\
2458922.622285 & --15.5 $\pm$ 6.2 & NIR \\
2459734.550026 & 19.6 $\pm$ 7.3 & NIR \\
2459748.431043 & 32.1 $\pm$ 4.5 & NIR \\
2459981.593216 & 27.2 $\pm$ 5.9 & NIR \\
2460070.581508 & 35.0 $\pm$ 6.0 & NIR \\
2460108.425963 & 8.0 $\pm$ 4.3 & NIR \\
2460157.360901 & 32.0 $\pm$ 5.4 & NIR \\
2460313.717922 & 25.2 $\pm$ 6.1 & NIR \\
2460360.666196 & 16.6 $\pm$ 6.1 & NIR \\
2460361.672895 & 39.6 $\pm$ 7.9 & NIR \\
\noalign{\smallskip}
\hline
\end{tabular}
\end{center}
\end{table}
\addtocounter{table}{-1}

\begin{table}[H]
\begin{center}
\caption[]{Continued.}
\begin{tabular}{rcc}
\hline
\hline
\noalign{\smallskip}
               &  RV             &              \\
BJD            &  m\,s$^{-1}$    & Instrument   \\
\noalign{\smallskip}
\hline                               
\noalign{\smallskip}
2460389.701010 & 18 $\pm$ 11 & NIR \\
2460492.433749 & --1.7 $\pm$ 5.5 & NIR \\
2460526.368382 & 42.2 $\pm$ 9.1 & NIR \\
2460638.729409 & 19.6 $\pm$ 7.3 & NIR \\
2460689.756309 & 13.9 $\pm$ 7.5 & NIR \\
2460731.661100 & 15.2 $\pm$ 5.3 & NIR \\
\noalign{\smallskip}
\hline
\end{tabular}
\end{center}
\end{table}

\section{Photometry time series}
\label{app:LCdata}
The nightly averaged photometric times series of GJ\,9773, and GJ\,508.2 used in this work are provided in Tables\,\ref{tab:LC_GJ9773} and \ref{tab:LC_GJ508}, respectively. The third column on each table indicates the observatory and filter used for each measurement.

\begin{table}[h]
\begin{center}
\caption[]{GJ\,9773 photometric time series.}
\label{tab:LC_GJ9773}
\begin{tabular}{rcc}
\hline
\hline
\noalign{\smallskip}
BJD            &  Norm. flux       & Instrument/Band   \\
\noalign{\smallskip}
\hline
\noalign{\smallskip}
2460420.661618 & 1.0031$\pm$0.0014 & TJO/R \\
2460423.653689 & 1.0044$\pm$0.0016 & TJO/R \\
2460425.648477 & 1.0007$\pm$0.0013 & TJO/R \\
2460433.646301 & 1.0015$\pm$0.0012 & TJO/R \\
2460437.636960 & 0.9993$\pm$0.0012 & TJO/R \\
2460442.626410 & 1.0042$\pm$0.0016 & TJO/R \\
2460443.623675 & 1.0104$\pm$0.0014 & TJO/R \\
2460449.619450 & 1.0054$\pm$0.0011 & TJO/R \\
2460451.607214 & 0.9977$\pm$0.0012 & TJO/R \\
2460456.603359 & 0.9993$\pm$0.0017 & TJO/R \\
\noalign{\smallskip}
\hline
\end{tabular}
\end{center}
\textbf{Notes:} Only a subset of the data analysed in this paper is shown here. A machine-readable version of the full dataset is available at CDS.
\end{table}

\begin{table}[h]
\begin{center}
\caption[]{GJ\,508.2 photometric time series.}
\label{tab:LC_GJ508}
\begin{tabular}{rcc}
\hline
\hline
\noalign{\smallskip}
BJD           &  Norm. flux   & Instrument/Band   \\
\noalign{\smallskip}
\hline              
\noalign{\smallskip}
2460244.710071 & 1.0066$\pm$0.0016 & TJO/R \\
2460249.711347 & 0.9978$\pm$0.0040 & TJO/R \\
2460255.722244 & 1.0028$\pm$0.0016 & TJO/R \\
2460256.722210 & 1.0024$\pm$0.0016 & TJO/R \\
2460261.720276 & 1.0030$\pm$0.0014 & TJO/R \\
2460263.720842 & 1.0056$\pm$0.0016 & TJO/R \\
2460265.720128 & 0.9976$\pm$0.0021 & TJO/R \\
2460266.735653 & 0.9977$\pm$0.0015 & TJO/R \\
2460267.652768 & 1.0067$\pm$0.0014 & TJO/R \\
2460269.718024 & 1.0028$\pm$0.0015 & TJO/R \\
\noalign{\smallskip}
\hline
\end{tabular}
\end{center}
\textbf{Notes:} Only a subset of the data analysed in this paper is shown here. A machine-readable version of the full dataset is available at CDS.
\end{table}

\section{GJ\,3512 exoplanet detection limits}
\label{app:astrometry}
\begin{figure}[h]
\centering
\includegraphics[width=\columnwidth]{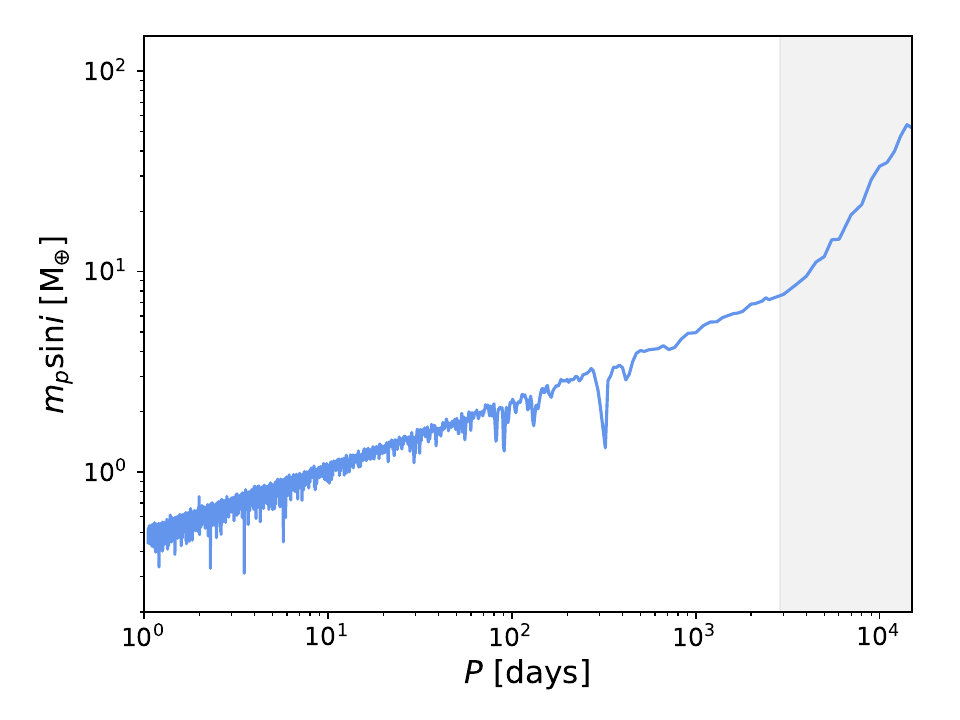}
\caption{Planetary mass detection limit for the CARMENES VIS RVs for GJ\,3512 after subtracting the two planet model in Table\,\ref{tab:RVfit}. Any planet above the blue line would be detected with a with 99\% confidence. The grey shaded region indicate the period values greater than the time baseline of the RV measurements.}
\label{fig:detection_limit_GJ3512}
\end{figure}

In order to evaluate whether we could detect additional planetary signals in the CARMENES VIS RVs of GJ\,3512, we subtracted the two planet model described in Sect.\,\ref{sec:GJ3512_rv} and we followed the procedure outlined in \cite{Bonfils2013}. We added a planetary signal to the residuals of the fit assuming a circular orbit, and we increased the semi-amplitude of this signal until it is detected with a 1\% FAP level by means of a GLS periodogram analysis. A total of 12 phase angles were tested for each period. Figure\,\ref{fig:detection_limit_GJ3512} illustrates the minimum planetary mass that would be detectable at such significance for each orbital period. Only planetary mass values above the depicted line would be found within the CARMENES VIS RVs. The grey shaded area depicts the long-period region beyond the time baseline of the CARMENES observations.

\section{Astrometric constraints}
\label{app:astrometry}
\begin{figure*}[h]
\centering
\includegraphics[width=0.4\textwidth]{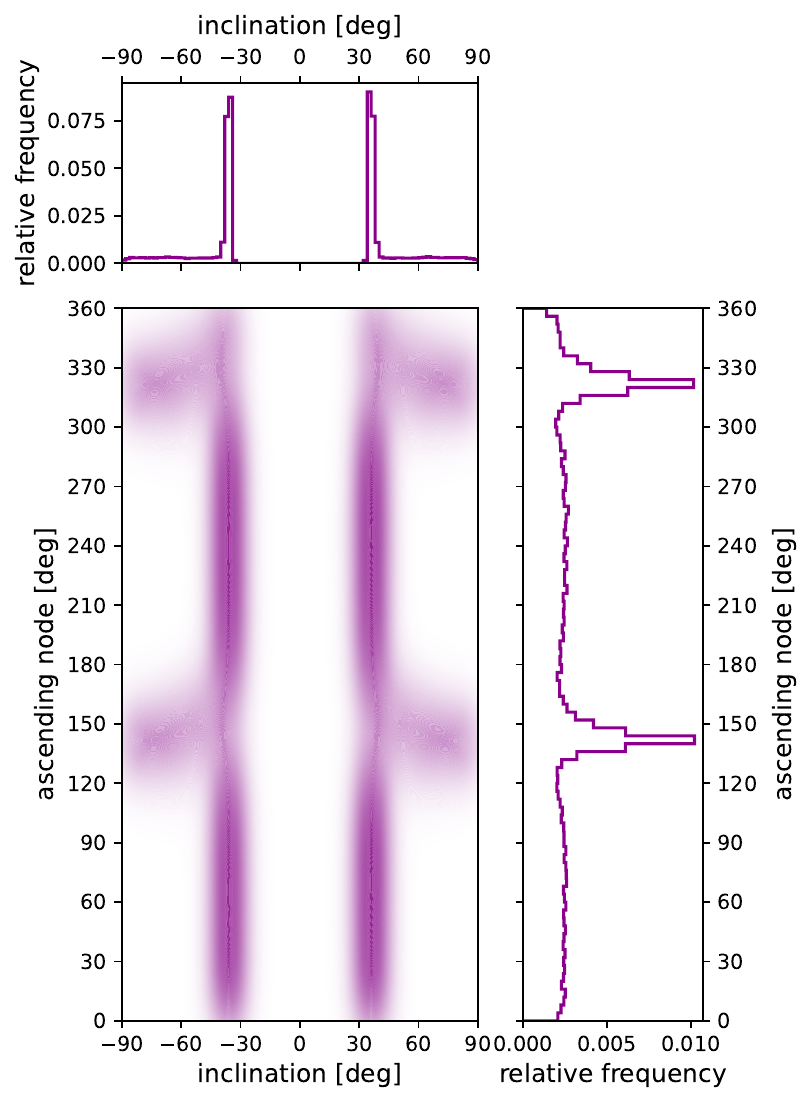}
\includegraphics[width=0.4\textwidth]{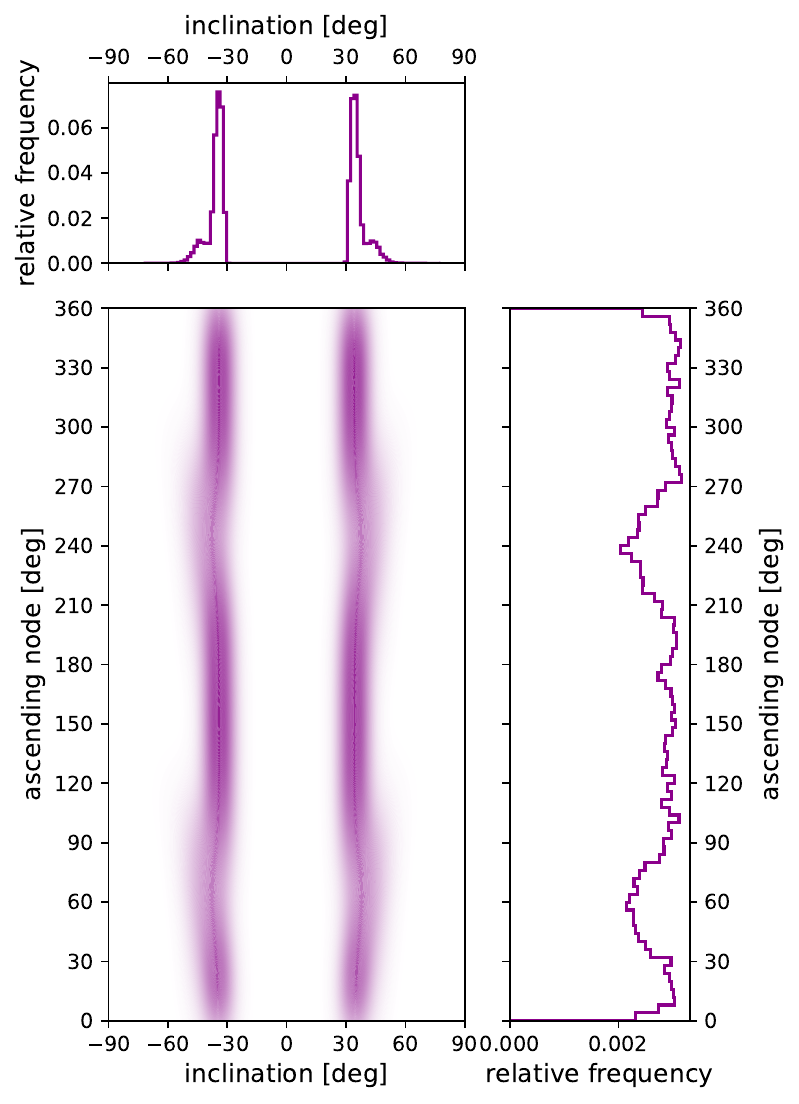}
\caption{Posterior distributions of the orbital inclination and ascending node derived from the orbital parameters fitted to the RV data and the \textit{Gaia} astrometric excess noise for the GJ\,3512 (left) and GJ\,317 (right) systems. In both cases, planets b and c are considered to have the same orbital inclination.}
\label{fig:astrometry_GJ3512_GJ317}
\end{figure*}

\begin{figure*}[h]
\centering
\includegraphics[width=0.4\textwidth]{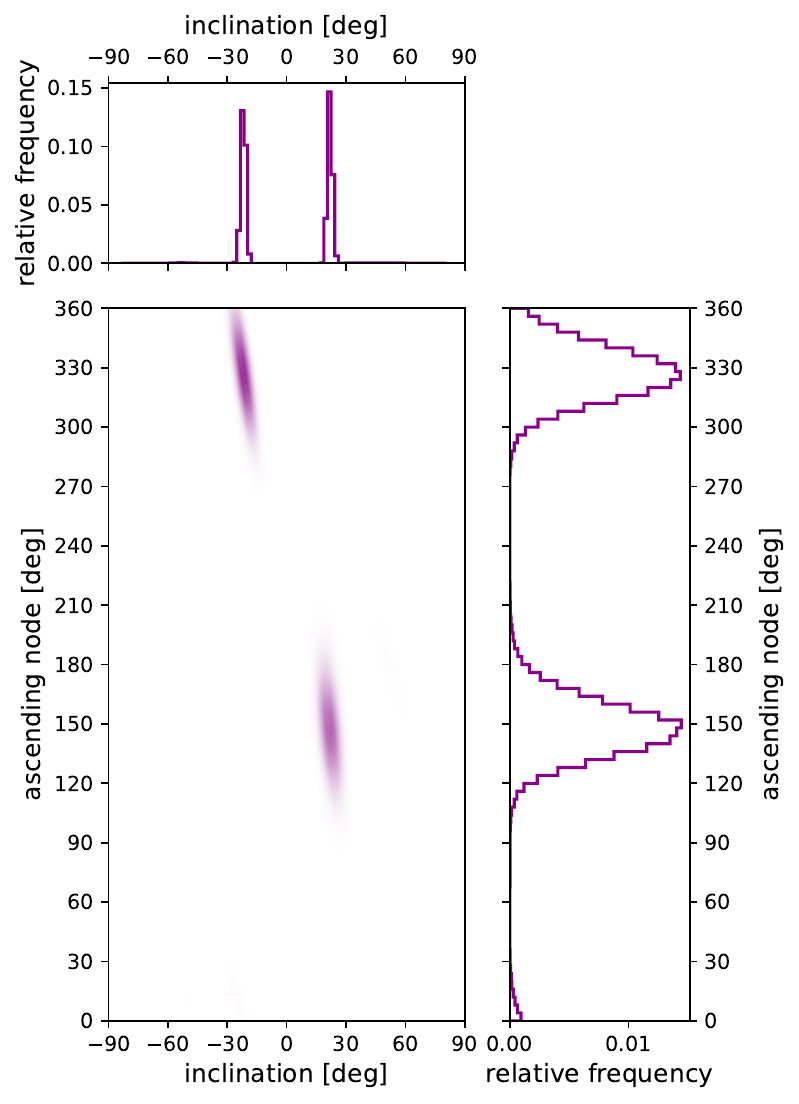}
\includegraphics[width=0.4\textwidth]{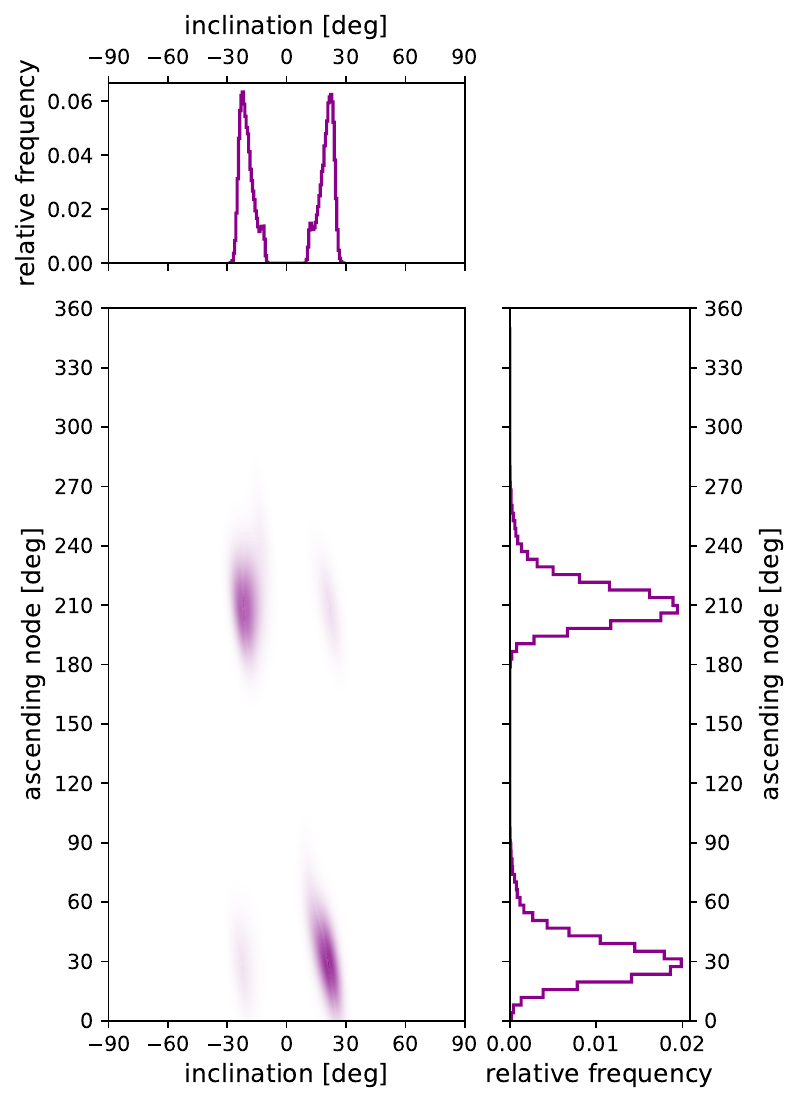}
\caption{Posterior distributions of the orbital inclination and ascending node derived from the orbital parameters fitted to the RV data, the \textit{Gaia} astrometric excess noise and the proper motion anomaly for the GJ\,463 (left) and  GJ\,9773 (right) systems.}
\label{fig:astrometry_GJ463_GJ9773}
\end{figure*}

\begin{figure}[h]
\centering
\includegraphics[width=0.4\textwidth]{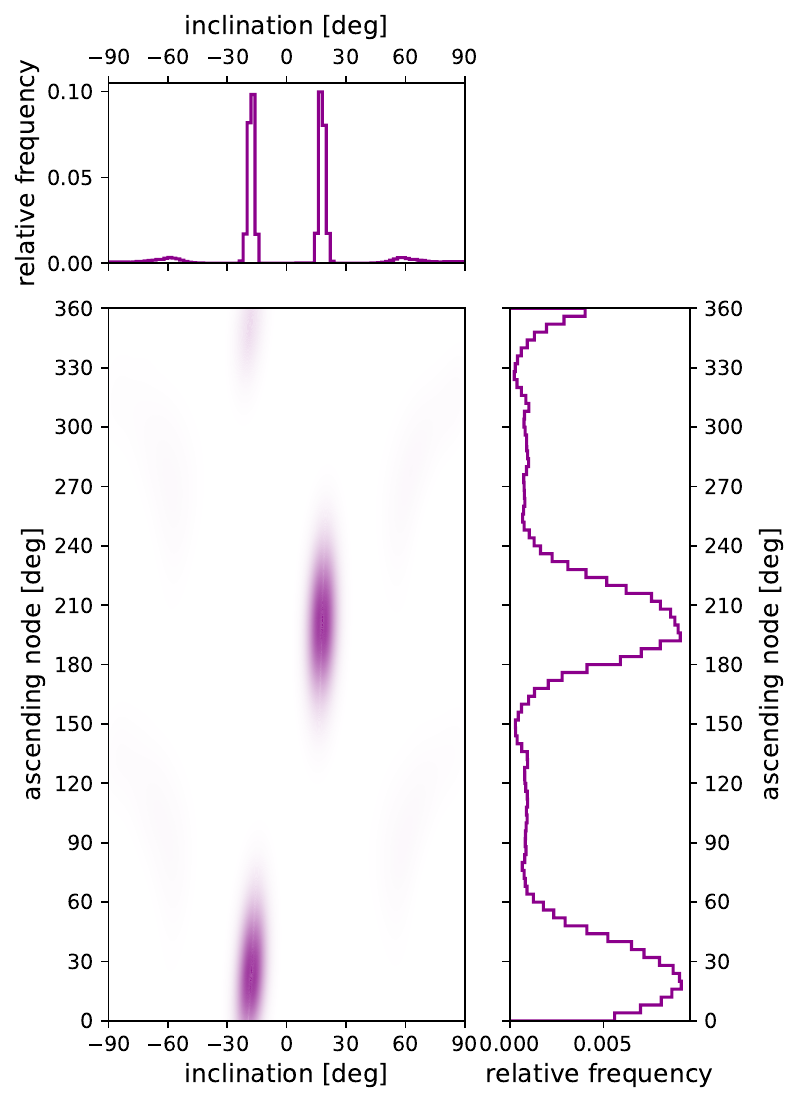}
\caption{Posterior distributions of the orbital inclination and ascending node derived from the orbital parameters fitted to the RV data, the \textit{Gaia} astrometric excess noise and the proper motion anomaly for the GJ\,508.2 system. The parameters fitted to the lower orbital period limit are used.}
\label{fig:astrometry_GJ508}
\end{figure}

Figures\,\ref{fig:astrometry_GJ3512_GJ317}, \ref{fig:astrometry_GJ463_GJ9773}, and \ref{fig:astrometry_GJ508} illustrate the distribution of the orbital inclination ($i$) and ascending node ($\Omega$) constrained from the \textit{Gaia} astrometric excess noise, the proper anomalies between \textit{Gaia} and Hipparcos, and the orbital parameters obtained from the RV analyses as described in Section\,\ref{sec:astrometry}. Table\,\ref{tab:astrometry} lists the parameters derived from these distributions.

\begin{table}[h]
\caption[]{Orbital parameters and absolute planetary masses constraints derived from the RV orbital fitting and the \textit{Gaia} astrometric excess noise and proper motion anomaly.}
\label{tab:astrometry}
\begin{tabular}{rcccc}
\hline
\hline
\noalign{\smallskip}
            & $i$                      & $\Omega$               & $m_{\rm b}$              & $m_{\rm c}$  \\
System      & (deg)                    & (deg)                  & (M$_{\rm Jup}$)       & (M$_{\rm Jup}$)\\
\hline
\noalign{\smallskip}
GJ 3512 bc  &   36.0\aunc{27}{-1.2}    & 142.5\aunc{7.2}{-108}  & 0.78\aunc{0.04}{-0.27} & 0.77\aunc{0.04}{-0.26}  \\[0.8ex]
            & --36.0\aunc{1.2}{-27}    & 322.5\aunc{6.5}{-107}  & 0.78\aunc{0.04}{-0.27} & 0.77\aunc{0.04}{-0.26}  \\[1.2ex]
GJ 317  bc  &   33.9\aunc{6.1}{-1.4}   & --                     & 3.03\aunc{0.13}{-0.41} & 2.56\aunc{0.14}{-0.35}   \\[0.8ex]
            & --34.1\aunc{1.5}{-6.0}   & --                     & 3.01\aunc{0.14}{-0.40} & 2.55\aunc{0.14}{-0.34}   \\[0.8ex]
\noalign{\smallskip}
\hline
\noalign{\smallskip}
GJ 463 b    &   21.8\aunc{1.5}{-1.1}   & 147\aunc{15}{-13}      & 4.44\aunc{0.25}{-0.29} & --  \\[0.8ex]
            & --21.8\aunc{1.4}{-1.2}   & 326\aunc{16}{-12}      & 4.44\aunc{0.31}{-0.24} & --  \\[1.2ex]
GJ 9773 b   &   22.3\aunc{1.3}{-6.1}   &  29.0\aunc{14}{-7.4}   & 1.43\aunc{0.52}{-0.08} & --  \\[0.8ex]
            & --22.3\aunc{6.1}{-1.3}   &  209.4\aunc{13}{-7.8}  & 1.43\aunc{0.52}{-0.08} & --  \\[1.2ex]
GJ 508.2 b  &   17.9\aunc{4.8}{-1.9}   & 199\aunc{21}{-27}      & 6.0\aunc{0.70}{-1.2} $^{(a)}$  & -- \\[0.8ex]
            & --17.8\aunc{1.9}{-4.8}   &  20\aunc{26}{-21}      & 6.0\aunc{0.70}{-1.2} $^{(a)}$  & --  \\[0.8ex]
\noalign{\smallskip}
\hline
\end{tabular}

\textbf{Notes:} $^{(a)}$ The lower minimum mass threshold of the planet GJ\,508.2 b ($m_{\rm b}\sin i = 1.85$\,\mjup) is used to estimate these value.
\end{table}

\end{appendix}
\end{document}